\documentclass[usenatbib]{mn2e}
\usepackage{graphicx}
\usepackage{color}
\title[Yunnan-III models for EPS]
{Yunnan-III models for Evolutionary population synthesis}

\author[F. Zhang, L. Li, Z. Han, et al.]
{F.~Zhang\thanks{E-mail: gssephd@public.km.yn.cn; zhang\_fh@hotmail.com}$^{1,2}$, L.~Li$^{1,2}$, Z.~Han$^{1,2}$, Y.~Zhuang$^{1,2,3}$ and X.~Kang$^{1,2}$\\
$^1$National Astronomical Observatories/Yunnan Observatory, Chinese Academy of Sciences, Kunming, 650011, China \\
$^2$Key Laboratory for the Structure and Evolution of Celestial Objects, Chinese Academy of Sciences, Kunming, 650011, China \\
$^3$Graduate University of the Chinese Academy of Science, Beijing 100049, China
}

\font\hf = cmsl7 scaled \magstep 0

\begin{document}

\date{\today}

\pagerange{\pageref{firstpage}--\pageref{lastpage}}

\pubyear{2005}

\maketitle

\label{firstpage}

\begin{abstract}
We build the Yunnan-III evolutionary population synthesis (EPS) models by using the {\hf MESA} stellar evolution code, BaSeL stellar spectra library and the initial mass functions (IMFs) of Kroupa and Salpeter, and present colours and integrated spectral energy distributions (ISEDs) of solar-metallicity stellar populations (SPs) in the range of 1Myr-15\,Gyr.
The main characteristic of the Yunnan-III EPS models is the usage of a set of self-consistent solar-metallicity stellar evolutionary tracks (the masses of stars are from 0.1 to 100M$_{\rm \odot}$). This set of tracks is obtained by using the state-of-the-art {\hf MESA} code.

{\hf MESA} code can evolve stellar models through thermally pulsing asymptotic giant branch (TP-AGB) phase for low- and intermediate-mass stars.
By comparisons, we confirm that the inclusion of TP-AGB stars make the $V-K$, $V-J$ and $V-R$ colours of SPs redder and the infrared flux larger at ages log($t$/yr)$\ga$7.6 (the differences reach the maximum at log($t$/yr)$\sim$8.6, $\sim$0.5-0.2\,mag for colours, $\sim$ 2 times for $K-$band flux).
We also find that the colour-evolution trends of Model with-TPAGB at intermediate and large ages are similar to those from the {\hf STARBURST99} code, which employs the Padova-AGB stellar library, BaSeL spectral library and the Kroupa IMF.
At last, we compare the colours with the other EPS models comprising TP-AGB stars (such as CB07, M05, V10 and POPSTAR), and find that the $B-V$ colour agrees with each other but the $V-K$ colour exists larger discrepancy among these EPS models ($\sim$1\,mag when 8$\la$log($t$/yr)$\la$9).

The stellar evolutionary tracks, isochrones, colours and ISEDs can be obtained on request from the first author or from our website (http://www1.ynao.ac.cn/$\sim$zhangfh/). Using the isochrones, you can build your EPS models. Now the format of stellar evolutionary tracks is the same as that in the {\hf STARBURST99} code, you can put them into the {\hf STARBURST99} code and get the SP's results.
Moreover, the colours involving other passbands or on other systems (for example, HST $F439W-F555W$ colour on AB system) can also be obtained on request.
\end{abstract}


\section{Introduction}
\label{Section:intro}

Evolutionary population synthesis (EPS) method was first introduced by \citet{tin68}, and mainly is used to study the unsolved stellar population (SP).
Key ingredients in EPS models are: (i) the library of evolutionary tracks used to calculate isochrones in the colour-magnitude diagram; (ii) the library of stellar spectra adopted to derive the integrated spectral energy distribution (ISED) or magnitudes and colours in suitable passbands; and (iii) the initial mass function (IMF) used to evaluate the relative proportions of stars in the various evolutionary phases.

Stellar evolutionary tracks and isochrones play an important role in EPS models.
However, many EPS models use the combination of several stellar evolutionary libraries [such as the models of \citet{wor94}, {\hf P\'EGASE} \citep{fio97,fio99}, etc.]. This implies that the basic stellar properties are discontinuous along the composite tracks \citep[][hereafter M08]{mar08}.
Moreover, due to theoretical and observational limitations, many stellar spectra libraries only cover the limited parameter (such as effective temperature $T_{\rm eff}$, surface gravity log$g$, metallicity [Fe/H], etc.) ranges. Therefore, some EPS models use the combination of several stellar spectra libraries.

Based on the above-mentioned facts, in this paper we will build the Yunnan-III EPS models and present the ISEDs and colours of solar-metallicity SPs by using a set of self-consistent stellar evolution models (computed from Modules for Experiments in Stellar Astrophysics [{\hf MESA}] code, its properties will be described in Section~\ref{Section:model-des}), a stellar spectra library (BaSeL-2.2, \citealt[][hereafter LCB97, LCB98]{lej97,lej98}) and the IMFs of \citet[][hereafter K01]{kro01} and \citet[][hereafter S55]{sal55}.
Before the construction of the Yunnan-III models, we have built two kinds of EPS models. The Yunnan-I models are constructed for SPs without binary interactions \citep{zha02} by using the rapid single stellar evolution algorithms \citep[][hereafter HPT00]{hur00}. The Yunnan-II models\footnote{http://www1.ynao.ac.cn/$\sim$zhangfh/} are constructed for both SPs with and without binary interactions \citep{zha04b,zha05a} by using Monte Carlo simulation and the rapid single and binary stellar evolution algorithms (HPT00, \citealt[][hereafter HTP02]{hur02}).

Another main reason of building the Yunnan-III models is that stars on thermally pulsing asymptotic giant branch (TP-AGB) phase are often neglected or added following individual recipes \citep[hereafter M05]{mar05} in EPS models. This is because computations of the TP-AGB phase by means of complete evolution code are very demanding in terms of computational time \citep[][hereafter MG07]{mar07}. So far, there exists several EPS models that have considered the contribution of TP-AGB stars. These models would be recalled in Section~\ref{Section:model-ovw}. According to the used technique of dealing with TP-AGB stars, these EPS models are divided into two groups.

With 'fuel consumption' technique, \citet[][hereafter M98]{mar98} and M05 have considered the contribution of TP-AGB stars in her EPS models. The advantages of the 'fuel consumption' approach are summarized by M05. Based on her models, she and her colleagues have investigated the redshift $z \sim 2$ passive and star-forming galaxies \citep{mar06,mar10} and luminous red galaxies \citep{mar09}.

With 'isochrone synthesis' technique, some models [such as the models of CB11 and their older CB07 version \citep{bru11}, Flexible Stellar Population Synthesis (hereafter FSPS, \citealt{con09,con10a,con10b}), \citet[][hereafter V10]{vaz10}, {\hf STARBURST99} \citep[hereafter {\hf SB99},][]{lei99,lei10,vaz05}-v6.02, POPSTAR \citep{mol09}, Yunnan-I and II \citep{zha02,zha05a}, etc.] have included TP-AGB stars.
In these models, the evolution of TP-AGB stars is obtained either by perturbation function or by a synthetic code.

The Yunnan-I and II EPS models have used TP-AGB evolution from perturbation function of the HPT00 rapid stellar evolution code. HPT00 claimed that this method can produce the core-mass ($M_{\rm c}$) and luminosity ($L$) relation of the third dredge-up (TDU).

Then, what is a TP-AGB synthetic code? In a TP-AGB synthetic code, the stellar evolution is described by means of simplified relations (such as $M_{\rm c}$-$L$ relation). It can be divided into detailed and simple codes.
(i) In a detailedly synthetic code, the simplified relations are derived from complete stellar models, and convective dredge-up and mass loss are tuned by means of a few adjustable parameters (MG07).
A very detailedly synthetic code was developed by \citet{mar96,mar98TP} and \citet[][and references therein]{mar01s}.
MG07 and their colleagues have used an explicitly synthetic code, in which the TDU and hot bottom burning (HBB) are considered, to present the evolution of TP-AGB stars and the first isochrones considering the TDU \citep{mar01}.
(ii) In a simply synthetic code, the TDU and the over-luminosity above the $M_{\rm c}$-$L$ relation caused by HBB are neglected. \citet[][hereafter BBCFN-1994]{ber94} and \citet[][hereafter G00]{gir00} have used a simply synthetic method to present the TP-AGB evolution (M08).

Among those 'isochrone synthesis' EPS models including TP-AGB stars by a synthetic code, the TP-AGB evolution of the CB11, CB07 and FSPS models is from an explicitly synthetic code, while that of the V10 and POPSTAR models is from a very simply synthetic code.
Looking at the ingredients of the above EPS models including TP-AGB stars by a synthetic code (refer to Section 3), we find that all the above-mentioned models used the TP-AGB evolutionary tracks from Padova group.
The reason of the above mentioned situation is that the use of the so-called synthetic code is the only viable alternative at that time in order to provide extended grids of TP-AGB models that reproduce basic observational constraints (MG07).

However, it is no doubt that stellar evolution model is another method of obtaining TP-AGB evolution. Stellar evolution models often parameterize physical processes, this is contrary to the synthetic method \citep{her05}.

In this work, building the Yunnan-III models with 'isochrone synthesis' technique, we will compute the stellar evolutionary tracks including TP-AGB phase by the {\hf MESA} code, which
can compute the evolution of low- and intermediate-mass stars from zero-age main sequence through TP-AGB phase. Therefore, the Yunnan-III EPS models can be compared with those EPS models that have considered TP-AGB stars, and can be an alternative one in galaxy studies.
It is worth noting that the TDU is considered in the {\hf MESA} star and massive TP-AGB stars show the expected HBB behavior, including the avoidance of the C-star phase for a 5\,M$_\odot$, $Z$=0.01 stellar model track despite efficient TDU \citep{pax11}.

The outline of the paper is as follows. In Section 2 we describe the Yunnan-III EPS models. In Section 3 we recall some EPS models considering TP-AGB stars. In Sections 4 and 5 we present the results (evolutionary tracks, isochrones, colours and ISEDs of SPs) and comparisons. Finally we present a summary and conclusions in Section 6.

\section{Model description}
\label{Section:model-des}
In this section, we describe the employed ingredients (including the {\hf MESA} stellar evolution code, BaSeL stellar spectra library and IMFs) in the Yunnan-III EPS models. The main characteristics are listed in Table~\ref{Tab:moe-rev}.

\subsection{{\hf MESA} stellar evolution code: input physics and parameters}
Stellar evolution models form the basic of EPS studies, and therefore we devote much of the remainder of this section to describing aspects of the {\hf MESA} models in details.

The {\hf MESA} stellar evolution code was built by \citet{pax11}, and began as an effort to improve upon the {\hf EZ} stellar evolution code \citep{egg71,pax04}. During the construction, its design and implementation were influenced by a number of stellar evolution and hydrodynamic codes (therefore, it can calculate the evolutions of massive stars).
{\hf MESA} is a state-of-the-art stellar evolution code, it uses the most updated equation of state (EOS, can handle the low mass and high degree of degeneracy), opacity, nuclear reaction rates, element diffusion data, and atmosphere boundary conditions.
Moreover, {\hf MESA} code employs modern software engineering tools and techniques to target modern computer architectures, and its numerical and computational methods employed allow it to consistently evolve stellar models through challenging phase,
for example, {\hf MESA} code allows non-stop computations through the helium core flash toward the end of the AGB evolution for low-mass stars \citep{pax11}.
At last, {\hf MESA} code uses the adaptive mesh refinement and sophisticated
time-step controls, therefore, it is able to solve the fully coupled structure and composition equations simultaneously.
In the following, we describe the input physics and parameters used by us in the {\hf MESA} code.

{\bf Network:} the 'basic' network of eight isotopes ($^1$H, $^3$He, $^4$He, $^{12}$C, $^{14}$N, $^{16}$O, $^{20}$Ne and $^{24}$Mg) is the default value.
We use the 21 isotope reaction network (approx21.net), inspired by the 19 isotopy network\footnote{similar to Frank Timmes' APPROX19, see MESA/data/net$\_$ data/nets/README} in \citet{wea78}, that is capable of efficiently generating accurate nuclear energy generation rates from hydrogen burning through silicon burning.
This network includes linkages for PP-I, steady-state CNO cycles, a standard $\alpha-$chain, heavy ion reactions, and aspects of photodisintegration in $^{54}$Fe (see Section 7.3 of \citealt{pax11}).

{\bf EOS:} we use the {\hf MESA}, HELM \citep{tim00} and PC \citep{pot10} EOS tables. The {\hf MESA} EOS table is constructed from the 2005 update of the OPAL \citep{rog02} and SCVH \citep{sau95} EOS tables.

{\bf Opacity:} we use the opacity of \citet[hereafter GS98]{gre98} instead of \citet[][hereafter GN93, by default]{gre93}, with the low temperature opacities taken from \citet{fer05} with metal ratios given by GS98. The low $T$ opacities of \citet{fer05} include the effects of molecules and grains on the radiative opacity.

{\bf Mixing length parameter:} for $M < 30 \rm M_{\odot}$, mixing length parameter $\alpha =2$; for $M > \rm 30 M_{\odot}$, $\alpha =1$.

{\bf Overshooting:} {\hf MESA} treats convective mixing as a diffusive process with a diffusion coefficient. The overshoot mixing diffusion coefficient
$D_{\rm OV} = D_{\rm conv,0} {\rm exp}(-{2z \over f\lambda_{P,0}})$,
in which $D_{\rm conv,0}$ is the mixing-length-theory derived diffusion coefficient at a user-defined location near the Schwarzschild boundary, $\lambda_{P,0}$ is the pressure scale height at that location, $z$ is the distance in the radiative layer away from that location, and $f$ is an adjustable parameter.
By default, the overshooting is turn off. We switch on it when star mass $M>1.8\,\rm M_{\odot}$, and turn off it when $M<1.1\,\rm M_{\odot}$. In the mass range of $1.1-1.8\,\rm M_{\odot}$, overshooting is gradually enabled. Moreover, convective overshooting is only allowed at locations with $m \ge 0.001\,M$ (by default).
At the upper and lower convective boundaries for H-burning and non-burning convective zones, the adjust parameter $f=0.01$.

{\bf Mass loss:} two kinds of mass loss are needed to be mentioned. The first one is the mass loss of red giant branch (RGB) and AGB stars. By default, there is no RGB and AGB wind. We use the \citet[][$\dot{M}_{\rm R} = 4 \times 10^{-13} \eta_{\rm R} LR/M, \eta_{\rm R}=0.5$]{rei75} and \citet[][$\dot{M}_{\rm B} =  4.83  \times 10^{-9} \eta_{\rm B}  L^{2.7} M^{-2.1} \dot{M}_{\rm R}/\eta_{\rm R}$, $\eta_{\rm B}=0.1$]{blo95} mass loss laws for RGB and AGB stars, respectively. The second one is the enhanced mass loss due to rotation. By default, it is considered and the coefficient is 0.43. In this work, we set it 0 for the reason of comparison with the other EPS models.

{\bf Atmosphere:} the default atmosphere is a simple photosphere model, i.e. do not integrate and just estimate for $\tau=2/3$. We use the default one.

{\bf Comments:} {\bf (i) the TDU:} in the {\hf MESA} code, the TDU is considered and massive TP-AGB stars show the HBB behavior.
{\bf (ii) the set of input parameters:} using the above set of input parameters and physics, we obtain the luminosity of 1\,M$_\odot$ star at an age of 4.57$\times$10$^9$yr, log($L/\rm L_\odot$)=$-$0.01 for the GS98 opacity while log($L/\rm L_\odot$)=0.008 for the GN93 opacity. These luminosities are within one hundredth of a magnitude.

\subsection{BaSeL stellar spectra library}
The BaSeL-2.2 stellar spectra library (LCB97, LCB98) is a semi-empirical one and provides an extensive and homogeneous grid of low-resolution theoretical flux distributions in the range of 9.1$-$160\,000\,nm and synthetic $UBVRIJHKLM$ colours for a large range of stellar parameters: 2000 $\le T_{\rm eff}$/K $\le$ 50\,000, $-1.02 \le$ log $g \le$ 5.50 and $-5.0 \le$ [Fe/H] $\le$ 1.0.
We adopt the BaSeL stellar spectra library for the following reasons.
(i) The BaSeL library has wider coverage of wavelength in comparison with other libraries (see \citealt{bru11}), so the EPS models based on this library can be used to the studies of galaxies at different redshift via multi-band SED fitting method.
(ii) The BaSeL library provides wider coverage of metallicity in comparison with the empirical stellar spectra libraries. Our next work is to give the EPS models at the other metallicities (even at zero metallicity). Therefore, we use this library for the sake of homogeneity.

At solar metallicity, the BaSeL library covers the region occupied by the TP-AGB stars on the log $g$-log$T_{\rm eff}$ plane, but does not include TP-AGB stars.
Therefore, we use the normal M-giant spectra, provided by the BaSeL library, for TP-AGB stars in this work.

The situation, whether or not the adopted stellar spectra library includes the spectra of TP-AGB stars, would affect significantly the infrared (IR) results of SPs comprising TP-AGB stars.
Recently, some EPS models have used the empirical library comprising TP-AGB stars (such as the M05 model).
Therefore, in the future, we will use these empirical library comprising TP-AGB stars in our models.

\subsection{IMF}
We use the IMF of K01:
\begin{equation}
\phi(M)_{_{\rm K01}} = \Biggl\{ \matrix {
          M^{-0.30}, & 0.01 \le M \le 0.08, \cr
          M^{-1.30}, & 0.08 \le M \le 0.50, \cr
          M^{-2.30}, & 0.50 \le M \le 100, \cr
          }
\label{eq.imfk01}
\end{equation}
where $M$ is the stellar mass in units of M$_\odot$.

In order to compare with the other EPS models, we also use the S55 IMF:
\begin{equation}
\phi(M)_{_{\rm S55}} = M^{-2.35},  0.1 \le M \le 100.,
\label{eq.imfs55}
\end{equation}
where $M$ is the stellar mass in units of M$_\odot$.

\section{Overview of EPS Models}
\label{Section:model-ovw}

\begin{table*}
\tiny
\centering
\caption{Main characteristics of EPS (M05, MS11, CB11, CB07, FSPS, V10, {\hf SB99}, POPSTAR, YUNNAN-I/II an III) models. '/' means that the corresponding EPS ingredient can be chosen. '+' means that the EPS ingredient is from several sources.}
\begin{tabular}{lll l}
\hline
    Name     & stellar evolution library/isochrone   & stellar spectra library         & IMF  \\
\hline

  M05        & Cassisi + Geneva + Padova   & BaSeL-3.1+LM02                   & S55/K01 \\
  MS11       &     -                       & Pickles/ELODIE/STELIB/MILES+LM02 & S55/K01/Cha03 \\
\hline
  CB07       &  Padova-1994/Padova-2000 + MG07                              & BaSeL-3.1/STELIB/Pickles     &  S55/Cha03 \\
  CB11       &  Padova-1994 + BGMN08 + MG07 + G10                           & MILES+other                    & Cha03 \\
\hline
  FSPS       &  Baraffe et al.(1998) + MG07 + M08                           & BaSeL-3.1+LM02                   & van Dokkum (2008)  \\
  V10        &   G96($Z$=$10^{-4}$) + G00($Z$$>$$10^{-4}$) isochrones       & MILES                            &  unimodal/bimodal/K01/K01-cor     \\
  {\hf SB99}       &  Padova-AGB/Padova-STD/Geneva-STD/Geneva-HIG           & choose   &  change \\
  POPSTAR    & Padova-1994+BGS98 & BaSeL+other    & S55/Ferrini/K01/Cha03 \\
  Yunnan-I/II& Yunnan-I/II       & BaSeL/HRES/BLUERED & MS79/S55 \\
  Yunnan-III & Yunnan-III        & BaSeL              & S55/K01\\
\hline
\end{tabular}
\label{Tab:moe-rev}
\end{table*}

In this section, we will overview those EPS models which have considered the contribution of TP-AGB stars. The main characteristics, including stellar evolution library, stellar spectra library and IMF, are summarized in Table~\ref{Tab:moe-rev}.

To distinguish which EPS models include TP-AGB evolution by an explicitly or a simply synthetic codes, we first give a brief description of the Padova stellar evolutionary tracks and isochrones.

(i) BBCFN-1994 ($Z$=0.0004, 0.004, 0.008, 0.02 and 0.05) and \citet[][hereafter G96, $Z$=$10^{-4}$]{gir96} presented the isochrones comprising TP-AGB evolution from a simply synthetic code. These two sets of isochrones are derived from the so-called Padova-1994 tracks, which are used in the \citet[][hereafter BC03]{bru03} and POPSTAR models. BBCFN-1994 isochrones are the basis of the following two sets of isochrones.

(ii) G00 used a simply synthetic prescription to include TP-AGB regime to the point of complete envelope ejection for stars within 0.15$\le$$M$/M$_\odot$$\le$7. In the work of G00, the updated opacities and EOS, and a moderate amount of convective overshoot are used, while TDU is not taken into account. Also, they built the isochrones ($Z$=0.0004, 0.001, 0.004, 0.008, 0.019 and 0.03, 10$^{7.8}$$-$$10^{10.2}$\,yr). This set of isochrones is used by the V10 models.
Further, combining with the work of BBCFN-1994, they built the Girardi-2002 isochrones (= [BBCFN-1994] + [G00] + [simplified TP-AGB]).

(iii) MG07 presented TP-AGB evolution by using a detailedly synthetic code and some recipes (semi-empirical prescriptions). They started from the physical conditions at the first thermal pulse from G00 and \citet{gir01} and computed the evolution up to the complete loss of the envelope via stellar winds (M08). In their models, several important theoretical improvements over previous calculations (such as, TDU and HBB) have been included.
Moreover, the prescriptions have been calibrated against near-IR data from the Large Magellanic Clouds (LMC) and Small Magellanic Clouds (SMC), including carbon star luminosity function and TP-AGB lifetime (star count) in MC clusters \citep{con09}.
M08 presented the Marigo-2008 isochrones based on the work of MG07 (= [BBCFN-1994] + [G00] + [MG07]). Based on the Marigo-2008 isochrones, \citet[][hereafter G10]{gir10} further built the isochrones with two kinds of corrections for low-mass and low-$Z$ AGB tracks.
The TP-AGB evolution of MG07 is used by the CB11, CB07, FSPS and {\hf SB99}-v6.02 models.

\subsection{M05 and the latest version MS11}
Maraston have used the 'fuel consumption theorem' to build a series of EPS models. In the M05 and the latest version \citep[][hereafter MS11]{mar11} EPS models, TP-AGB stars are semi-empirically included and calibrated with observational data.

In the M05 models, several stellar evolutionary libraries\footnote{Cassisi + Geneva for $10^{-3} \le Z \le$ 0.04 (Geneva for $10^{-3} \le t \le$ 30\,Myr); Cassisi for $Z=10^{-4}$ and Padova for $Z$=0.07. In the above descriptions, 'Cassisi' means the work of \citet{cas97a}, \citet{cas97b} and \citet{cas00}, 'Geneva' includes the work of \citet{sch92} and \citet{mey94}.},
the BaSeL-3.1 stellar spectra library (LCB97; LCB98; \citealt{wes02}), and the S55 ($\phi(M)_{\rm S55} = M^{-\alpha}, \alpha=2.35$) and K01 IMFs were used. For the spectra of TP-AGB stars, the empirical library by \citet[][hereafter LM02]{lan02} was used.

The MS11 models were built on the basis of M05 models by replacing the BaSeL-3.1 library by four empirical high-to-intermediate spectral resolution libraries, namely \citet{pic98}, ELODIE \citep{pru07}, STELIB \citep{leB03} and MILES \citep{san06}. In addition to the above updation, they also presented the results based on the IMF of \citet[][hereafter Cha03]{cha03}.

In the M05 models, six metallicities ($Z$=$10^{-4}$, $10^{-3}$, 0.01, 0.02, 0.04 and 0.07) are included in the age range from 3 $\times 10^{-6}$\,Gyr to 15\,Gyr (67 ages, 16 ages within 1$-$15\,Gyr for $Z$=$10^{-4}$ and 0.07).

In the MS11 models, several metallicities and ages are included. For different stellar spectra library, the metallicity and age coverage is different\footnote{
For Pickles: $Z$=0.02; 2.5Myr$-$12/15Gyr.
For STELIB: $Z$=0.01, 0.02 and 0.04; 200/30/400Myr$-$12/15Gyr.
For MILES: $Z$ = $10^{-4}$, $10^{-3}$, 0.01, 0.02 and 0.04; 5G/2G/55M/$\sim$6M(3G)/100Myr $-$15Gyr.
For ELODIE: $Z=10^{-4}, 10^{-3}$, 0.02 and 0.04; 6G/55M/ 3M/100Myr $-$ 11/15Gyr.
}.

\subsection{CB11 and their preliminary version CB07}
Charlot \& Bruzual have built a series of EPS models. In the CB11 and their preliminary version CB07 models, TP-AGB stars have been included.

The main characteristics of the CB11 models are as follows. They were built by using
(i) the  evolutionary tracks of \citet[][hereafter BGMN08]{ber08}\footnote{
In fact, CB11 used the tracks up to 15\,M$_\odot$ from the models with updated input physics by BGMN08. For stars more massive than 15\,M$_\odot$, in the range from 20 to 120\,M$_\odot$, CB11 used the so-called Padova-1994 tracks. The TP-AGB evolution of low- and intermediate-mass stars is followed according to the semi-empirical prescription of G10.}, the TP-AGB prescriptions of MG07 and G10,
(ii) the MILES stellar spectra library, complemented at the hot effective temperature end by theoretical model atmospheres from \citet{lan03}, \citet{martins05}, \citet{rod05} and \citet{lan07}, and at the cool end by the IRTF library \citep{ray09} and the models of \citet{ari09},
and (iii) the Cha03 IMF with the mass limits of 0.1 and 100\,M$_{\odot}$.
Here, it is emphasized that BGMN08 also presented TP-AGB tracks, based on the MG07 prescriptions but extrapolated to different chemical compositions of the stellar envelope \citep{bru11}.

The CB07 models are similar to the BC03 models, except the use of the MG07 prescriptions to describe TP-AGB evolution. Moreover, CB07 models complemented the results at $Z$=0.1.
The CB07 models used (i) two sets of stellar evolutionary tracks (Padova-1994 \& Padova-2000), (ii) two forms of IMF (S55 with $\alpha=-$2.35 \& Cha03), and (iii) high- (STELIB \& Pickles) and low-resolution (BaSeL-3.1) stellar spectra libraries. For both IMFs, the lower and upper mass limits are 0.1 and 100.\,$\rm M_\odot$.

The CB11 models have not been released, while the BC07 models are available. For the CB07 models, seven metallicities ($Z$=$10^{-4}$, 4x$10^{-4}$, 4x$10^{-3}$, 8x$10^{-3}$, 0.02, 0.05 and 0.1) and 220 ages (from $10^{5}$ to $10^{10.3}$\,yr) are included.

\subsection{FSPS:}
FSPS models also have considered the contribution of TP-AGB stars. These models make use of
(i) the non-evolving stellar models of \citet{bar98} for stars in the mass range 0.10 $ < M/{\rm M_\odot} < $ 0.15 and the models of MG07 and M08,
(ii) the BaSeL-3.1 library, complemented by the average TP-AGB spectra compiled by LM02 from more than 100 optical/near-IR spectra presented in \citet{lan00} (this is similar to that of M05),
and (iii) the IMF form advocated by \citet{van08}:
\begin{equation}
\tiny
\psi = {\rm dn \over {\rm d ln}M} = \bigl\{
\matrix{
A_{\rm l} (0.5n_{\rm c} m_{\rm c} )^{-x} {\rm exp} [{−({\rm log} M - {\rm log} m_{\rm c} )^2 \over 2\sigma^2} ],  (M \le n_{\rm c} m_{\rm c} ), \cr
A_{\rm h} M^{-x}, \ \ \ \ \ \ \ \ \ \ \ \ \ \ \ \ \ \ \ \ \ \ \ \ \ \ \ \ \ \ \ \ \ \ \ \ \ \ (M > n_{\rm c} m_{\rm c} ), \cr
}
\label{Eq:IMF-v08}
\end{equation}
in which A$_{\rm l}$ = 0.140, $n_{\rm c}$ = 25, $x$ = 1.3, $\sigma$ = 0.69 and A$_{\rm h}$ = 0.158. Variation of the IMF is incorporated in the characteristic mass m$_{\rm c}$. The lower and upper mass limits of 0.1 and 100\,M$_\odot$ are used, respectively.

\subsection{V10:}
Vazdekis also has built some EPS models. His latest models, V10, make use of
(i) the updated version of the models published in G96 for $Z$=$10^{-4}$ (the calculations are now compatible with G00) and the solar-scaled theoretical G00 isochrones for $Z > 10^{-4}$,
(ii) the MILES stellar spectra library,
and (iii) four forms of IMF: unimodal, bimodal, K01-universal and K01-revised (hereafter K01-cor). The slope of IMF for unimodal and bimodal forms is in the range of 0.3$-$3.3. The lower and upper mass limits are 0.1 and 100\,M$_\odot$, respectively.

V10 models provide the results at seven metallicities ($Z$= $10^{-4}$, 4x$10^{-4}$, $10^{-3}$, 4x$10^{-3}$, 8x$10^{-3}$, 0.019 and 0.03), the age range is different at different $Z$:
10$-$18\,Gyr at $Z$=$10^{-4}$ (only for IMF slopes $\le$ 1.8);
0.07$-$18\,Gyr at $Z$=4x$10^{-4}$; and
0.06$-$18\,Gyr at $10^{-3} \le Z \le 0.03$.

\subsection{\hf SB99 v6.02:}
{\hf SB99} models were built by Leitherer and his colleagues. We use the 6.02 version. The description of the input physics is given by \citet{lei99}, \citet{vaz05} and \citet{lei10}. The main characteristics are as follows.
\begin{itemize}
\item
Four sets of stellar evolutionary tracks are provided, with each set including five metallicities. These four sets are the Geneva tracks with standard and high mass-loss rates (hereafter Geneva-STD \& Geneva-HIG), the Padova tracks as updated by G00 (hereafter Padova-STD), and the Padova tracks with the inclusion of TP-AGB stars (hereafter Padova-AGB) following the prescription of \citet{vas93}.
For Geneva tracks, $Z$=$10^{-3}$, 0.004, 0.008, 0.02 and 0.04. For Padova tracks, $Z$= 0.0004, 0.004, 0.008, 0.02 and 0.05.
\item
Five sets of stellar atmosphere libraries are provided to choose. BaSeL library is one of five.
\item
By default, the Kroupa IMF with two mass intervals (the exponent $\alpha=[-1.3, -2.3]$, the mass boundary $M_{\rm {cut}}$ = [0.1, 0.5, 100.]\,$\rm M_{\odot}$) is used.
The IMF form can be changed according to the need.
\end{itemize}

\subsection{POPSTAR:}
The POPSTAR models were built by \citet{mol09}, and the ingredients are described as follows.

\begin{itemize}
\item POPSTAR models adopt stellar evolutionary tracks from \citet{bre93}, \citet{fag94a,fag94b} and G96 (i.e. the Padova-1994 tracks), and combine a revision of the Padova \citep[][hereafter BGS98]{bre98} isochrones used in \citet{gar98}.
\item POPSTAR models mainly use the BaSeL stellar atmosphere models, which are for normal stars.
    For O, B and Wolf-Raylet (WR) stars, POPSTAR models take the non-local thermodynamic equilibrium (NLTE) blanketed models by \citet{smi02}  for $Z$ = 0.001, 0.004, 0.008, 0.02 and 0.04.
    For post-AGB and planetary-nebula stars, POPSTAR models use the \citet{rau03} NLTE models, covering the temperature range of 50\,000-190\,000 K and log$g$ of 5.00-8.00.
    For higher temperatures, POPSTAR models use blackbodies.
    These models include all elements from H to Ni, and they are available for two values of metallicities: $Z$ = 0.002 and 0.02.
    POPSTAR models use the first metallicity spectra for their three metal-poor ($Z \le$ 0.004) isochrones, and the solar abundance spectra for the three other metallicity ($Z \ge 0.008$) isochrones.
\item POPSTAR models include six IMFs. They are the S55 law with $\alpha = -2.35$ in the range of 0.85-120, 0.15-100 and 1.00-100\,M$_\odot$; the \citet{fer90}, K01 and Cha03 IMFs with masses between 0.15 and 100\,M$_\odot$.
\end{itemize}

At last, POPSTAR models provide the results at six metallicities ($10^{-4}$, 4x$10^{-4}$, 4x$10^{-3}$, 8x$10^{-3}$, 0.02 and 0.05) and $\sim$106 ages ($10^{5.}-10^{10.18}$yr).

\subsection{Yunnan-I and II:}
In the Yunnan-I and II models, rapid single and binary stellar evolution codes (HPT00, HTP02) are used. In the HPT00 and HTP02 codes, TP-AGB stars are obtained by perturbation function.
In the Yunnan-I models, only for SPs without binary interactions, the BaSeL-2.2 and several IMFs [S55 with different $\alpha$ and \citet[][hereafter K93]{kro93}] are used.
In the Yunnan-II models, for both SPs with and without binary interactions, several [including the BaSeL-2.2, BLUERED \citep[][$\sim0.3\,\rm\AA$]{ber08}, and HRES \citep[][$\sim0.1\,\rm\AA$]{gon05}] stellar spectra libraries are used.
For SPs with binary interactions, both IMFs for the primaries and secondaries are needed. We use the IMFs of S55 with $\alpha=-2.35$ and \citet{mil79} for the primaries. The lower and upper mass limits are 0.1 and 100\,M$_\odot$, respectively.

Both models can provide the results at seven metallicities ($Z$=$10^{-4}$, 3x$10^{-4}$, $10^{-3}$, 4x$10^{-3}$, 0.01, 0.02 and 0.03) and in the range from $10^5$ to $10^{10.18}$yr.

\section{Results and comparisons with {\sl SB99} models}
\label{Section:result}
\begin{figure}
\includegraphics[angle=270,scale=.37]{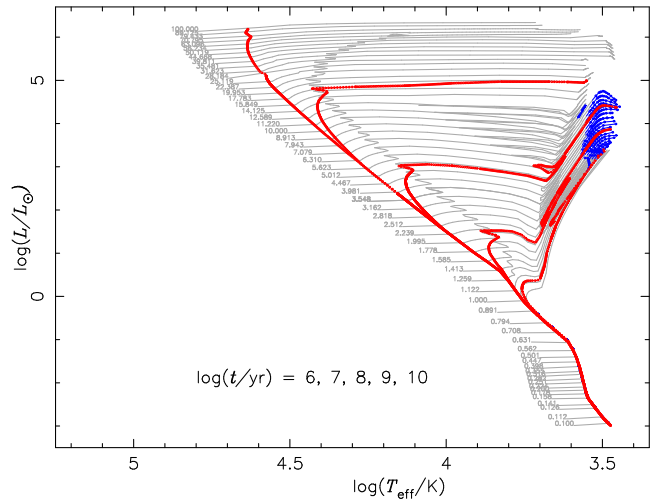}
\caption{Stellar evolutionary tracks (grey line) and isochrones (red line). The masses of stars are in the range of 0.1$-100\,\rm M_{\rm \odot}$ and at a logarithmic interval of 0.05. The ages of isochrones are log($t$/yr) = 6, 7, 8, 9 and 10. The TP-AGB tracks are represented by blue symbols.}
\label{Fig:trks}
\end{figure}

\begin{figure}
\includegraphics[bb=11 17 666 521,angle=  0,scale=.37]{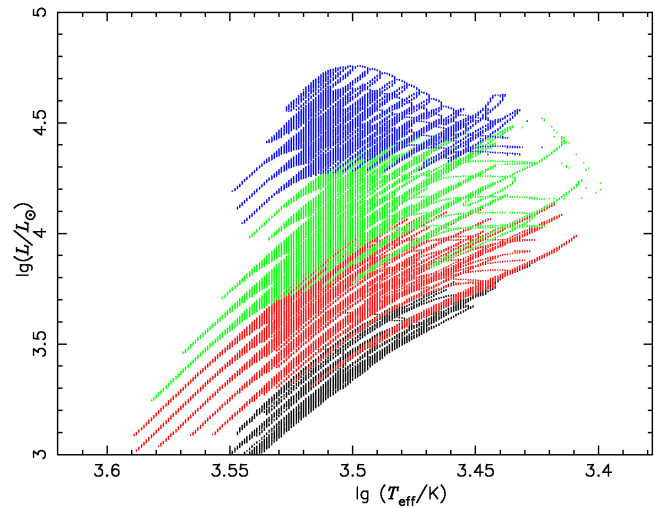}
\caption{TP-AGB phase on the HR diagram for stars with 1.0 $\le M/{\rm M_\odot} \le$ 7.9 [log($M{\rm /M_\odot})=(i-1)*0.05$, $i$=1-20. Black points are for $i$=1-5 (1.0-1.58${\rm M_\odot})$, red are for $i$=6-10 (1.78-2.82${\rm M_\odot})$, green for $i$=11-15 (3.16-5.01${\rm M_\odot})$ and blue for $i$=16-19 (5.62-7.94${\rm M_\odot})$].}
\label{Fig:tp-hrd}
\end{figure}

\begin{figure}
\includegraphics[angle=  0,scale=.400]{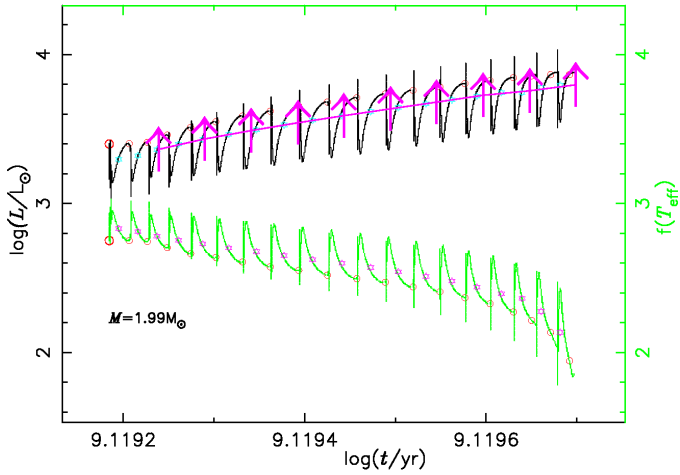}
\caption{Evolutions of luminosity $L$ (black line, left y-axis) and temperature $T_{\rm eff}$ [green line, right y-axis, $f(T_{\rm eff})$ = (log$T_{\rm eff}$-3.2)*8.] for a $M=1.99\,{\rm M_\odot}$ star on the TP-AGB phase. Also shown are the average $L'$ and $T'_{\rm eff}$ within a pulse (sixangles) and the interpolated points (uparrows, only for luminosity).}
\label{Fig:trk-tp}
\end{figure}

\begin{figure*}
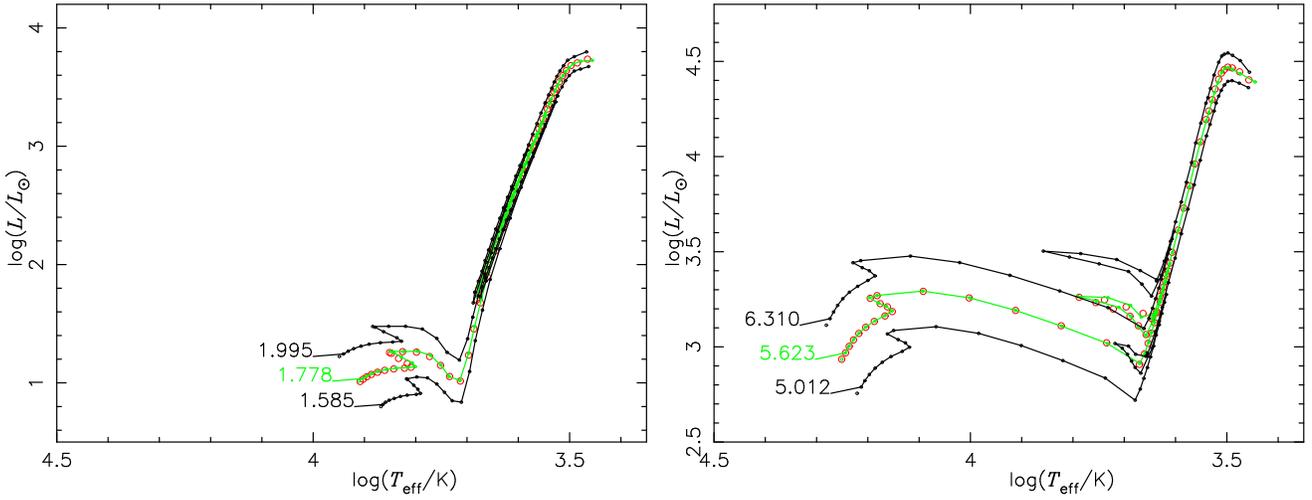

\includegraphics[bb=107 80 550 663,clip,angle=270,scale=0.420]{trk2itp-1.7.ps}
\includegraphics[bb=107 80 550 663,clip,angle=270,scale=0.420]{trk2itp-5.6.ps}
\caption{Comparisons between the model (green line) and interpolated (red open circles) tracks for a low-mass star ($M=1.78$\,M$\rm _\odot$, left panel) and an intermediate-mass star ($M=5.62$\,M$\rm _\odot$, right panel). In each panel, the tracks of the two adjacent stars (black line) are also presented.}
\label{Fig:trk-itp}
\end{figure*}

\begin{figure}
\centering
\includegraphics[angle=  0,scale=.350]{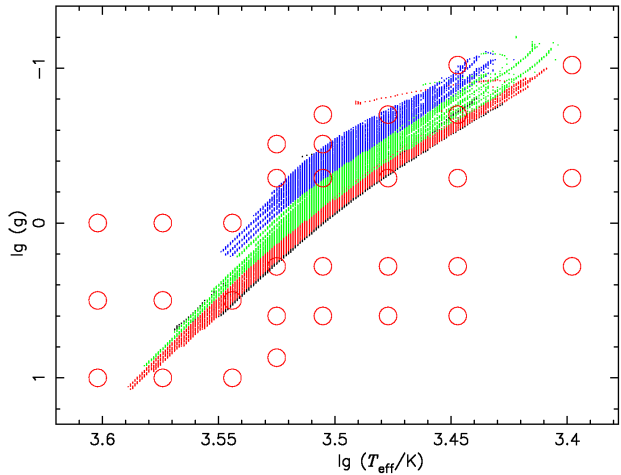}
\caption{Evolution of TP-AGB stars on the lg ($T_{\rm eff}$).vs.lg ($g$) plane. Also shown are the grid covered by the BaSeL stellar spectra library at solar metallicity.}
\label{Fig:tp-Teff-g}
\end{figure}

\begin{figure*}
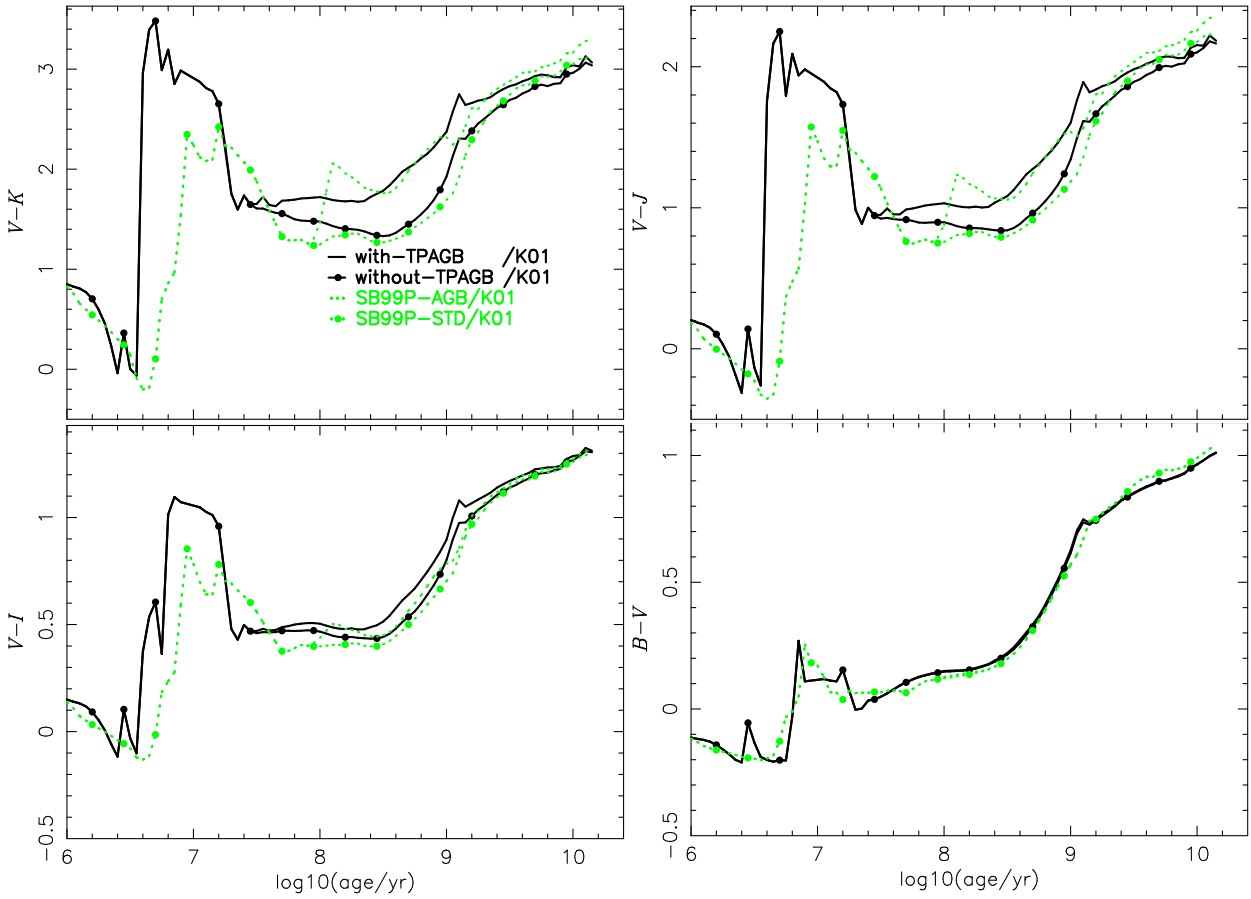

\includegraphics[bb=79 72 504 700,clip,angle=270,scale=.37]{VK-t3.ps}
\includegraphics[bb=79 72 504 700,clip,angle=270,scale=.37]{VJ-t3.ps}\\
\includegraphics[bb=79 72 560 700,clip,angle=270,scale=.37]{VI-t3.ps}
\includegraphics[bb=82 72 560 700,clip,angle=270,scale=.37]{BV-t3.ps}
\caption{Evolutions of $V-K$, $V-J$, $V-I$ and $B-V$ colours with the K01 IMF for Models with- (black, line) and without-TPAGB (black, line+solid circles). Also shown are the results with the K01 IMF from the {\hf SB99} code, which employs the Padova-AGB (Model SB99P-AGB, green, line) and Padova-STD (Model SB99P-STD, green, line+solid circles) evolutionary libraries, respectively.}
\label{Fig:ssps-color}
\end{figure*}

\begin{figure}
\includegraphics[bb=69 70 559 728,clip,angle=270,scale=.370]{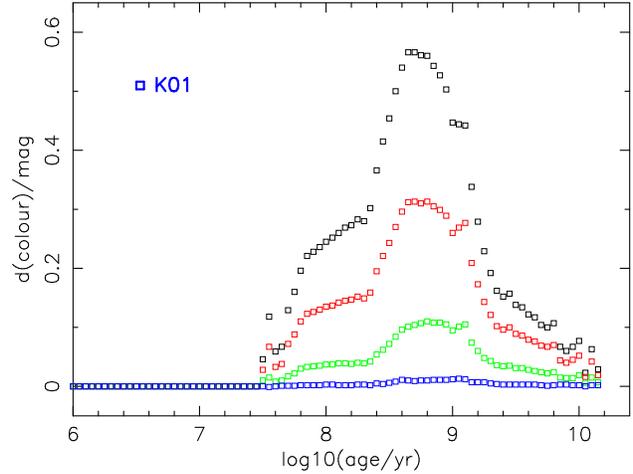}
\caption{The differences in the $V-K$, $V-J$, $V-I$ and $B-V$ colours (from top to bottom) between Models with-TPAGB and without-TPAGB for the K01 (rectangles) IMF.}
\label{Fig:dcolor}
\end{figure}

\begin{figure}
\includegraphics[bb=69 70 559 728,clip,angle=270,scale=.370]{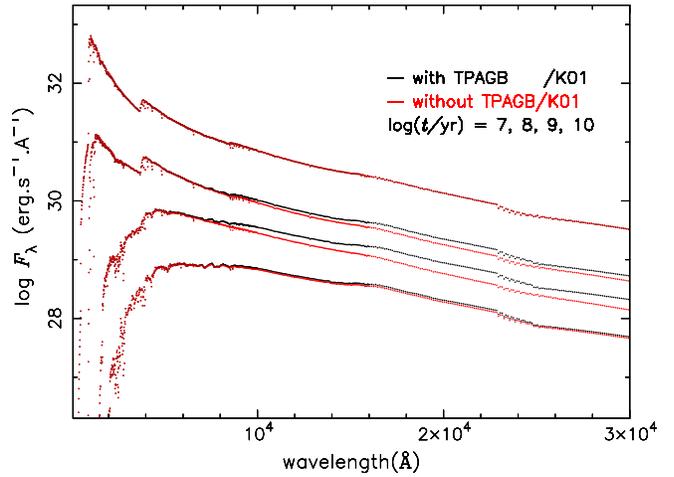}
\caption{The fluxes for the with-TPAGB-K01 (black line) and without-TPAGB-K01 models at ages of log($t$/yr) = 7, 8, 9 and 10.}
\label{Fig:ssps-ISEDs}
\end{figure}

\begin{figure}
\includegraphics[bb=69 70 559 726,clip,angle=270,scale=.37]{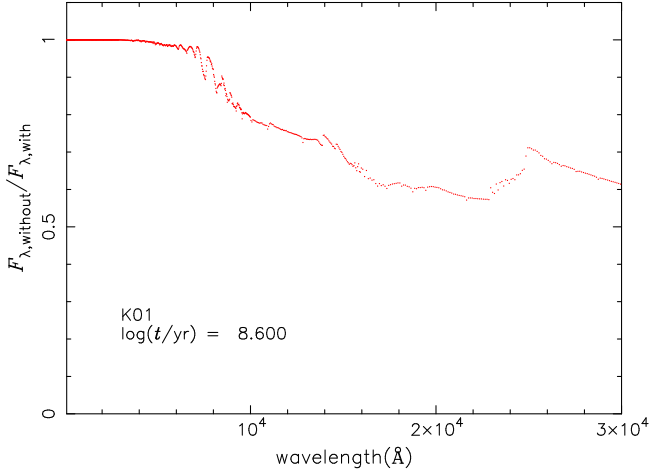}
\caption{The ratio of flux between the without-TPAGB-K01 ($F_{\lambda, \rm without}$) and with-TPAGB-K01 ($F_{\lambda, \rm with}$) models at an age of log($t$/yr) = 8.6.}
\label{Fig:ssps-ISEDs-rt}
\end{figure}

\begin{figure*}
\includegraphics[bb=82 26 580 712,clip=0,angle=270,scale=.350]{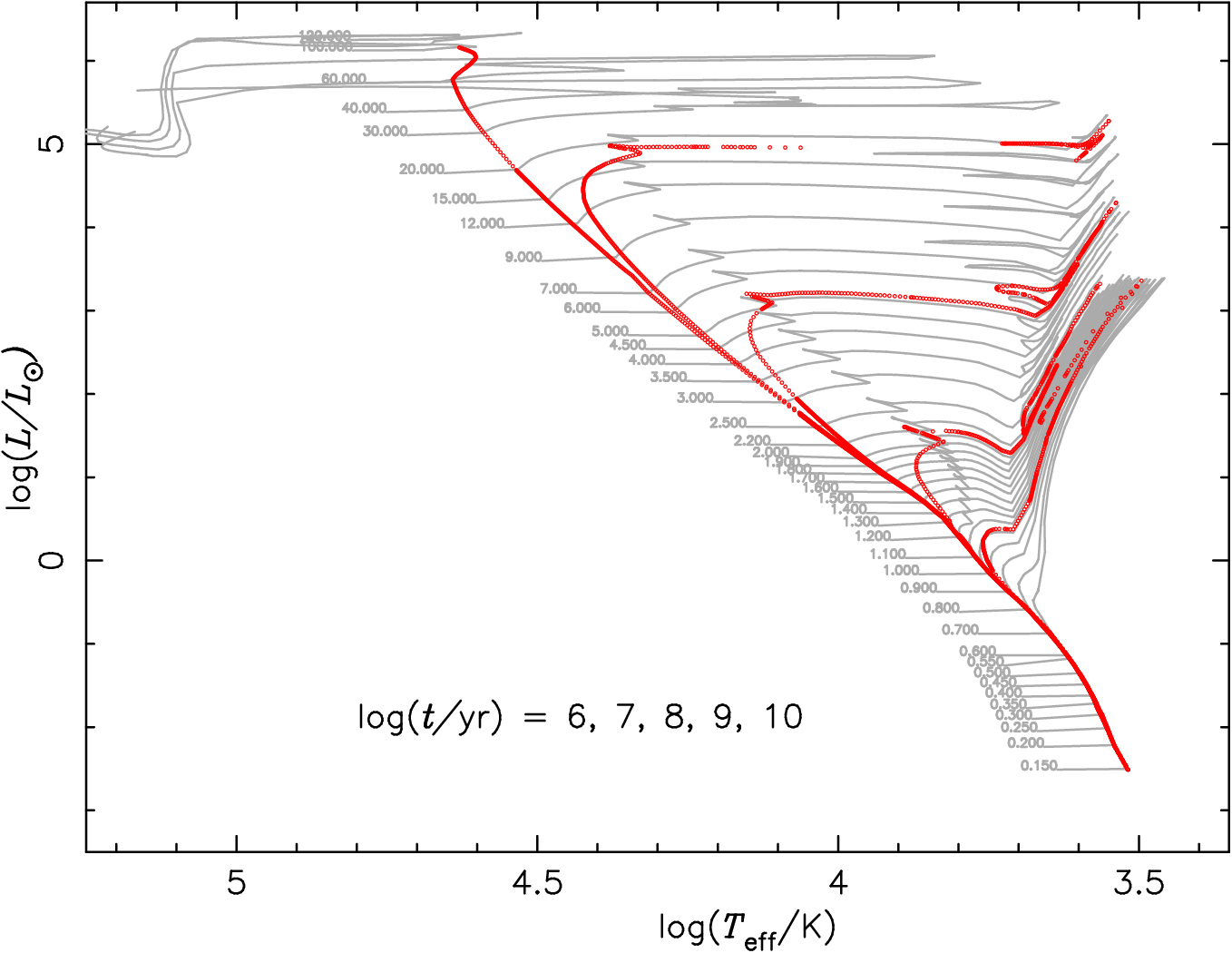}
\includegraphics[bb=82 26 580 712,clip=0,angle=270,scale=.350]{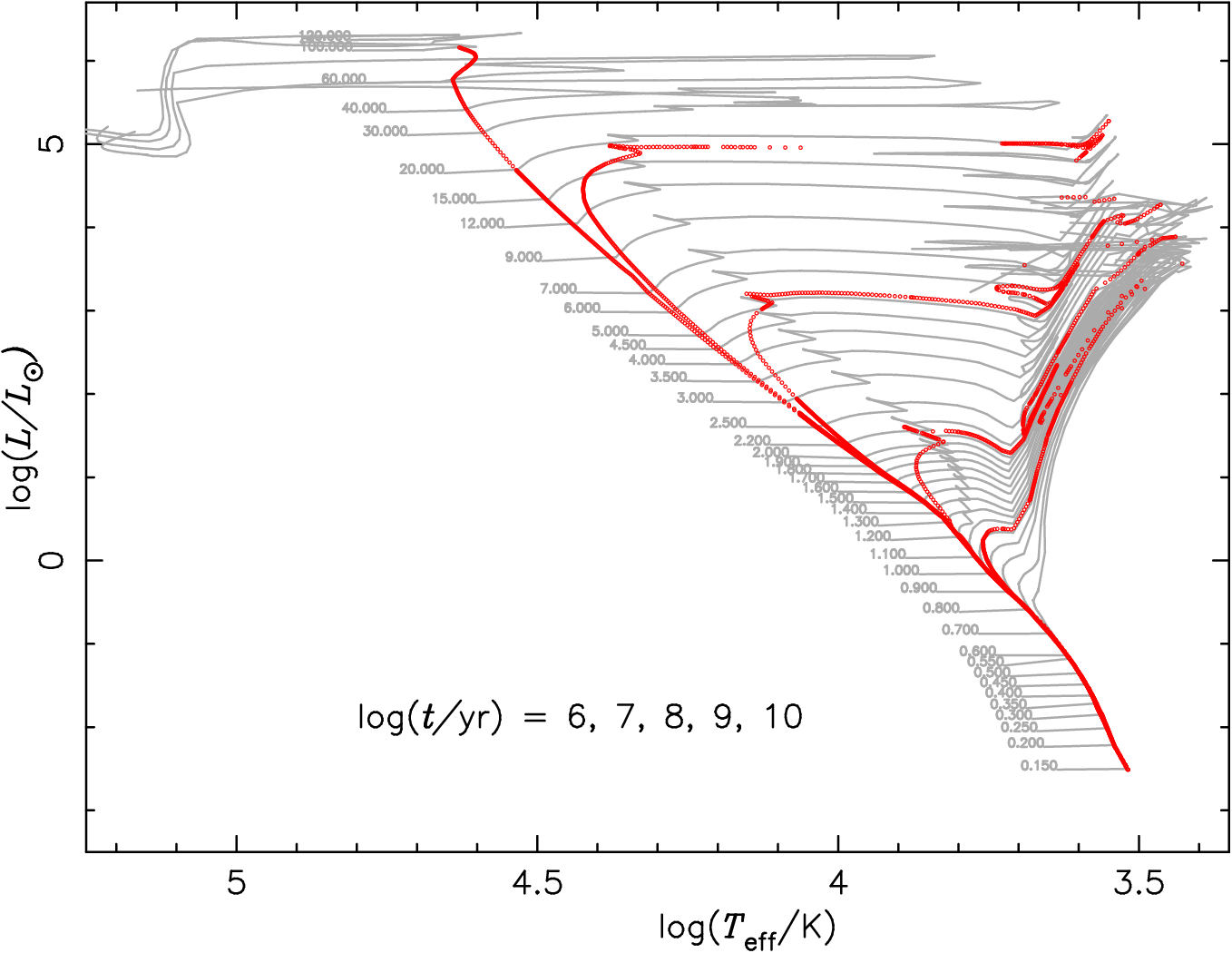}
\caption{Solar-metallicity stellar evolutionary tracks (grey line) and isochrones (red line) in the {\hf SB99 models}. The tracks and isochrones (i.e. SB99P-STD and SB99P-AGB) based on the Padova-STD and Padova-AGB libraries are presented in the left and right panels, respectively. The masses of stars are in the range of 0.15-120\,${\rm M_\odot}$. The ages of isochrones are log($t$/yr)=6, 7, 8, 9 and 10.}
\label{Fig:trks-other}
\end{figure*}

\begin{figure*}
\includegraphics[bb=69 70 551 662,clip,angle=270,scale=.40]{iso-com3.ps}
\includegraphics[bb=69 70 551 662,clip,angle=270,scale=.40]{iso-com3-small.ps}
\caption{Left panel is for the comparison in the solar-metallicity isochrone among different authors at ages of log($t$/yr)=6, 7, 8, 9 and 10. Black solid, red dashed and green dot-dashed lines are for the BBCFN-1994, Girardi-2000 and Marigo-2008 [without log($t$/yr)=6] isochrones, respectively.
Blue dot-dot-dot-dashed line, blue triangles and grey dotted line are for the SB99P-AGB, SB99P-STD and Yunnan-III isochrones.
For the sake of clarity, in the right panel we enlarge the low-$T_{\rm eff}$ and high-$L$ region (the region surrounded by the grey lines in the left panel). In the right panel, the tracks beyond TP-AGB phase have been neglected for the BBCFN-1994, Girardi-2000, Marigo-2008 and SB99P-AGB isochrones.
}
\label{Fig:iso-com}
\end{figure*}

In this section, we will present the colours and ISEDs for SPs. The ages are in the range of 1Myr-15Gyr. To investigate the effect of IMF, we perform two sets of calculations: one uses the K01 IMF whereas another uses the S55 IMF. The magnitudes and colours for the models with the K01 and S55 IMFs are presented in Tables~\ref{Tab:rsts-A} and~\ref{Tab:rsts-B}, respectively. The ISEDs and  isochrones can be obtained on request from the first author or from our website (http://www1.ynao.ac.cn/$\sim$zhangfh/). Moreover, in Tables~\ref{Tab:rsts-A} and~\ref{Tab:rsts-B} we only give several Vega colours on Johnson-Cousin system, the colours involving other passbands (such as $b-y$ colour) or those on AB system can also be obtained on request.

At last, we will give the comparisons in the colours between our and the {\hf SB99} models. The comparisons with the other literatures will be given in the next section. The reason, we put the comparisons with the {\hf SB99} models in this section, is that the {\hf SB99} code comprises two sets of stellar evolutionary tracks: Padova-STD and Padova-AGB (comprising and neglecting TP-AGB stars). This can be used to compare the effect of TP-AGB stars on the results between ours and theirs.

\subsection{Stellar evolutionary tracks and their simplification}

\subsubsection{Stellar evolutionary tracks}
We use the {\hf MESA} version 3709 and the set of input physics and parameters described in Section~\ref{Section:model-des} to calculate stellar evolution models for a set of solar-metallicity stars (i.e. hydrogen abundance $X$=0.70 and metallicity $Z$=0.02). This set of stellar evolution models, which are presented by grey lines in Fig.~\ref{Fig:trks}, forms the stellar evolutionary library of the Yunnan-III EPS models. The masses of stars are from 0.1 to $100 \,\rm M_{\rm \odot}$ and at a logarithmic interval of $\Delta$log($M/{\rm M_\odot})=0.05$. Using these tracks, we can build the isochrones of SPs.

TP-AGB stars are obtained by the {\hf MESA} stellar evolution code in the Yunnan-III EPS models. This is quiet different from the methods of using a synthetic code or perturbation function, by which TP-AGB evolution is obtained and used by the other EPS models. Moreover, TP-AGB stars give more contribution to the IR flux. Therefore, we give some detailed descriptions of the results concerning TP-AGB stars and discussions in Appendix~\ref{Section:appen-TPAGB}.

\subsubsection{Simplification}
In order to conveniently interpolate evolutionary tracks and build isochrones (see below), we simplify the stellar evolutionary tracks on TP-AGB phase, which is shown in Fig.~\ref{Fig:tp-hrd} [gives the TP-AGB tracks for stars with 1.0 $\le M/{\rm M_\odot} \le$ 7.9 on the Hertzsprung-Russell (HR) diagram], for intermediate- and low-mass stars:
we firstly obtain the average time $t'$, luminosity $L'$ and effective temperature $T'_{\rm eff}$ within each pulse by using $X'={\int X {\rm d}t \over \int \rm dt} (X= L$, $T_{\rm eff}$ or $t$), then obtain the simplified tracks by interpolating between these average points.
In Fig.~\ref{Fig:trk-tp}, we illustrate this process as an example of a $M=1.99\,\rm M_\odot$ star: the blue and purple sixangles denote the average $L'$ and $T'_{\rm eff}$ at epoch $t'$, the purple uparrows are those simplified values by interpolation.
In Fig.~\ref{Fig:trks}, the blue points are those simplified tracks on TP-AGB phase for different stars.

\subsection{Interpolation of evolutionary tracks and construction of isochrones}
One of the important steps in EPS models is to interpolate stellar evolutionary tracks by using those tracks computed by the {\hf MESA} code. This step will directly affect the results of EPS models.
To check this step, we will compare the interpolated with model tracks of stars.
First, we choose an intermediate-mass star ($M$=5.62\,$\rm M_\odot$) and a low-mass star ($M$=1.78\,$\rm M_\odot$) from the evolutionary library,
then obtain their interpolated tracks by using their own two adjacent model tracks.
In Fig.~\ref{Fig:trk-itp}, we give the comparisons between the interpolated and model tracks. From it, we see that the interpolated tracks are in good agreement with the model ones on all phases (including the loop) for both intermediate- and low-mass stars.

Using these model and interpolated evolutionary tracks, we build the isochrones of SPs. The ages of SPs are in the range of 1\,Myr-15\,Gyr and at a logarithmic age interval of $\Delta$ log($t \rm /yr$) = 0.05. In Fig.~\ref{Fig:trks}, the constructed isochrones are also presented at ages of log($t \rm /yr$) = 6, 7, 8, 9 and 10.

\subsection{Colours and ISEDs of SPs}
Combining with BaSeL-2.2 stellar spectra library and the K01 IMF, we present the colours and ISEDs of solar-metallicity SPs in the range of 1Myr-15\,Gyr at a logarithmic age interval of 0.05 (84 ages).
The S55 IMF is the most commonly used IMF, nearly all EPS models present the results via the S55 IMF. Therefore, we also present the results for models with the S55 IMF.
As said in Section 2.2, the BaSeL library covers the region spanned by TP-AGB stars on the log\,$T_{\rm eff}$-log\,$g$ plane at solar metallcity (see Fig.~\ref{Fig:tp-Teff-g}), but does not comprise TP-AGB stars. That is to say, we use the normal M-giant spectra for TP-AGB stars.
In Fig.~\ref{Fig:tp-Teff-g}, we present the evolution of TP-AGB stars computed via the {\hf MESA} code and the grid covered by the BaSeL library at solar metallicity on the log\,$T_{\rm eff}$-log\,$g$ plane.

In Fig.~\ref{Fig:ssps-color}, we present the evolutions of $V-K$, $V-J$, $V-I$ and $B-V$ colours for models with the K01 IMF.
Because that the effect of IMF on the results is small (the differences in the colours between the adoption of the S55 and K01 IMFs are less than one tenth of a magnitude), this conclusion also has been made by M98 by using several IMFs, in Fig.~\ref{Fig:ssps-color} the results with the S55 IMF are not been given.
As said above, one main characteristic of the Yunnan-III EPS models is that the employed {\hf MESA} code can evolve stellar models through TP-AGB phase for intermediate- and low-mass stars. Therefore, in Fig.~\ref{Fig:ssps-color} we also present the results when neglecting TP-AGB stars (i.e. neglecting the blue points in Fig.~\ref{Fig:trks}).
We refer to the results for the SPs with and without TP-AGB stars as Models with-TPAGB and without-TPAGB, respectively. To distinguish the IMF, the name of IMF is the supplement to the name of model.

From Fig.~\ref{Fig:ssps-color}, we see that the inclusion of TP-AGB stars can make $V-K$, $V-J$ and $V-R$ colours of SPs redder at log($t$/yr)$\ga$7.6, as known from M98 and M05, and the longer is the passband-wavelength of one magnitude in a colour, the more significant is the discrepancy in the colour [for example, d$(V-K) >$ d$(V-I)$].
The discrepancy in the $B-V$ colour of SPs is insignificant at all ages.
The maximal differences in the $V-K$, $V-J$ and $V-R$ colours of SPs are at log($t$/yr)$\sim$8.6
(see Fig.~\ref{Fig:dcolor}). In Fig.~\ref{Fig:dcolor}, we present the differences in the $V-K$, $V-J$, $V-I$ and $B-V$ colours between Models with-TPAGB and without-TPAGB for the K01 IMF.
The reason for the above fact is that some stars along the isochrone are on TP-AGB phase at log($t$/yr)$\ga$7.6 (see Fig.~\ref{Fig:trks}, corresponding to the onset time of TP-AGB phase for a $\sim$8\,M$_\odot$ star [see Table~\ref{Tab:tpagb} and Fig.~\ref{Fig:during-tp}]). These TP-AGB stars are cooler, and would affect the IR flux but almost do not affect the visible flux.

In Fig.~\ref{Fig:ssps-ISEDs}, we present the fluxes ($F_\lambda$) for the with-TPAGB-K01 and without-TPAGB-K01 models at ages of log($t$/yr) = 7, 8, 9 and 10. From it, we see that the inclusion of TP-AGB stars would affect significantly the IR flux at ages of ${\rm log} (t{\rm /yr)}$ = 8.0 and 9.0.
For the sake of clarity, in Fig.~\ref{Fig:ssps-ISEDs-rt} we present the ratio of flux between the without-TPAGB-K01 and with-TPAGB-K01 models ($F_{\lambda,{\rm without}}/F_{\lambda,{\rm with}}$) at an age of log($t$/yr)=8.6. From it, we see that $F_{\lambda,\rm without}$ is $\sim$\,50 percent of $F_{\lambda,\rm with}$ at wavelength $\lambda \sim 2\,\mu$m (i.e. K-band) for a SP at an age of log($t$/yr) $\sim$ 8.6.

\subsection {Comparisons in colours with {\hf SB99} models:}
\label{Section:result-com-st99}
We also obtain the results from the {\hf SB99} code, which employs the same stellar spectra library (BaSeL spectral library) and IMFs (i.e. the S55 with $\alpha=-2.35$ and K01 IMFs) as in this work but different stellar evolution library. The {\hf SB99} code only uses the solar-metallicity Padova-AGB and Padova-STD libraries (i.e. the 44th and 34th tracks, respectively) in this work.
The adopted ingredients of the {\hf SB99} code are listed in Table~\ref{Tab:moe-rev-used}. We refer to the {\hf SB99} models employing the Padova-AGB and Padova-STD libraries as Models SB99P-AGB and SB99P-STD, respectively. The properties of Padova-STD and Padova-AGB tracks have been described in Section~\ref{Section:model-ovw}. The upper and lower mass limits of IMFs are 100. and 0.1\,M$_{\odot}$.
For the {\hf SB99}-v6.02 code, we choose such a set of input parameters and physics for the following reasons: discuss the sole effect of stellar evolution and do not introduce the effects of stellar spectra library and IMF on the results.

In Fig.~\ref{Fig:ssps-color} we also give the results with the K01 IMF from the {\hf SB99} code, do not present the results with the S55 IMF from the {\hf SB99} code because that the effect of IMF on the colours of Models SB99P-AGB and SB99P-STD is small.
In the following, the results from the {\hf SB99 code} and Yunnan-III models, unless otherwise indicated, mean those with the K01 IMF. From Fig.~\ref{Fig:ssps-color}, we see that the $V-K$, $V-J$ and $V-I$ colours of Models with-TPAGB/without-TPAGB are very different from those from the {\hf SB99} code for 6.6$\la$ log($t$/yr)$\la$7.6. This is caused by the large discrepancy in the stellar evolutionary tracks of massive stars. When 6.6$\la$log($t$/yr)$\la$7.6, the IR fluxes of SPs mainly are contributed by massive stars ($\ga$8\,M$_\odot$ at solar metallicity in this work), which evolutionary results are prone to be affected by the input parameters. In Fig.~\ref{Fig:trks-other}, we present the stellar evolutionary tracks and isochrones of the {\hf SB99} models. The isochrones are obtained with the help of the {\hf SB99} code and built on the basis of the Padova-STD and Padova-AGB libraries, respectively. From it, we see that the massive stars in the Padova-AGB and Padova-STD libraries evolve through WR phase, and do not extend to very cool region in comparison with those in our stellar evolution library (see Fig.\,1).

In general, at log($t$/yr)$\ga$7.6, the colour-evolution trends of Models with-TPAGB/without-TPAGB agree to Models SB99P-AGB/SB99P-STD, respectively.
That is to say, the isochrones of Models with-TPAGB/without-TPAGB coincide with those of Models SB99P-AGB/SB99P-STD, respectively.
In Fig.~\ref{Fig:iso-com}, we give the Yunnan-III, SB99P-AGB and SB99P-STD isochrones at ages of log($t$/yr) = 6, 7, 8, 9 and 10. From them, we indeed see that the Yunnan-III isochrones agree to the SB99P-AGB/SB99P-STD isochrones.
To be exactly, at log($t$/yr)$\ga$7.6, there exist differences in the colours between Models with-TPAGB and SB99P-AGB, and those between Models without-TPAGB and SB99P-STD. They are as follows.

(i) The colours (except $V-I$) of Models with-TPAGB/without-TPAGB are smaller than the corresponding ones of Models SB99P-AGB/SB99P-STD by less than 0.1\,mag at large ages [log($t$/yr)$\ga$9.0].

(ii) The colour evolution of Model with-TPAGB is smoother than that of Model SB99P-AGB in the range 7.6 $\la$ ${\rm log} (t{\rm /yr)}$ $\la$ 8.5, there is a jump in the $V-K$, $V-J$ and $V-I$ colours for Model SB99P-AGB at age log($t$/yr)$\sim$8.0. Within 7.6$\la$log($t$/yr)$\la$8.0 the colours of Model with-TPAGB are larger ($\sim$0.2-0.4\,mag), but within 8.0$\la$log($t$/yr)$\la$8.5 are smaller ($\sim$0.2-0.4\,mag) than those of Model SB99P-AGB.
All above phenomena are partly due to the difference in the maximal mass of stars experiencing TP-AGB phase $M_{\rm max,TPAGB}$. In the Padova-AGB library, $M_{\rm max,TPAGB}$ is 5\,M$_\odot$, while in our stellar evolution library it is greater than 8\,$\rm M_{\odot}$ (refer to Table~\ref{Tab:tpagb} and Fig.~\ref{Fig:during-tp} in Appendix~\ref{Section:appen-TPAGB-1}) at solar metallicity.
Therefore, the jump in the colours of Model SB99P-AGB appears at log($t$/yr) $\sim$ 8.0, this epoch corresponds to the TP-AGB phase of a $M=M_{\rm max,TPAGB}$ (i.e. 5\,M$_\odot$) star in the Padova-AGB library (see the 51$^{\rm th}$ line for a $Z$=0.02 and 5\,M$_\odot$ star in the Padova-AGB library from the {\hf SB99} code).
In the range of 7.6$\la$${\rm log} (t{\rm /yr)}$$\la$8.0, there are no TP-AGB stars along the SB99P-AGB isochrones, but there are TP-AGB stars along the Yunnan-III isochrones. Therefore, the colours of Model with-TPAGB are redder than those of Model SB99P-AGB.
In the range of 8.0$\la$${\rm log} (t{\rm /yr)}$$\la$8.5, there are TP-AGB stars along the SB99P-AGB isochrones, and their TP-AGB stars are cooler and redder than those on the Yunnan-III isochrones (see Figs.~\ref{Fig:trks-other} and~\ref{Fig:iso-com}, only $8.0 \le {\rm log} (t{\rm /yr)} \le 9.0$). Therefore, the colours of Model with-TPAGB are bluer than those of Model SB99P-AGB.

(iii) The contribution of TP-AGB stars to the IR flux and colours peaks at log($t$/yr)=8.6 for the Yunnan-III models. This epoch corresponds to the TP-AGB phase for a $Z$=0.02, $M\sim$3.16M$_\odot$ star (refer to Table~\ref{Tab:tpagb} in Appendix~\ref{Section:appen-TPAGB-1}). However, the maximal contribution of TP-AGB stars is at log($t$/yr)=8.0 for the {\hf SB99} models with the Padova tracks, this corresponds to the epoch of jump in the colours.

\section{Comparisons with the other EPS models}
\label{Section:result-oth}

\begin{figure*}
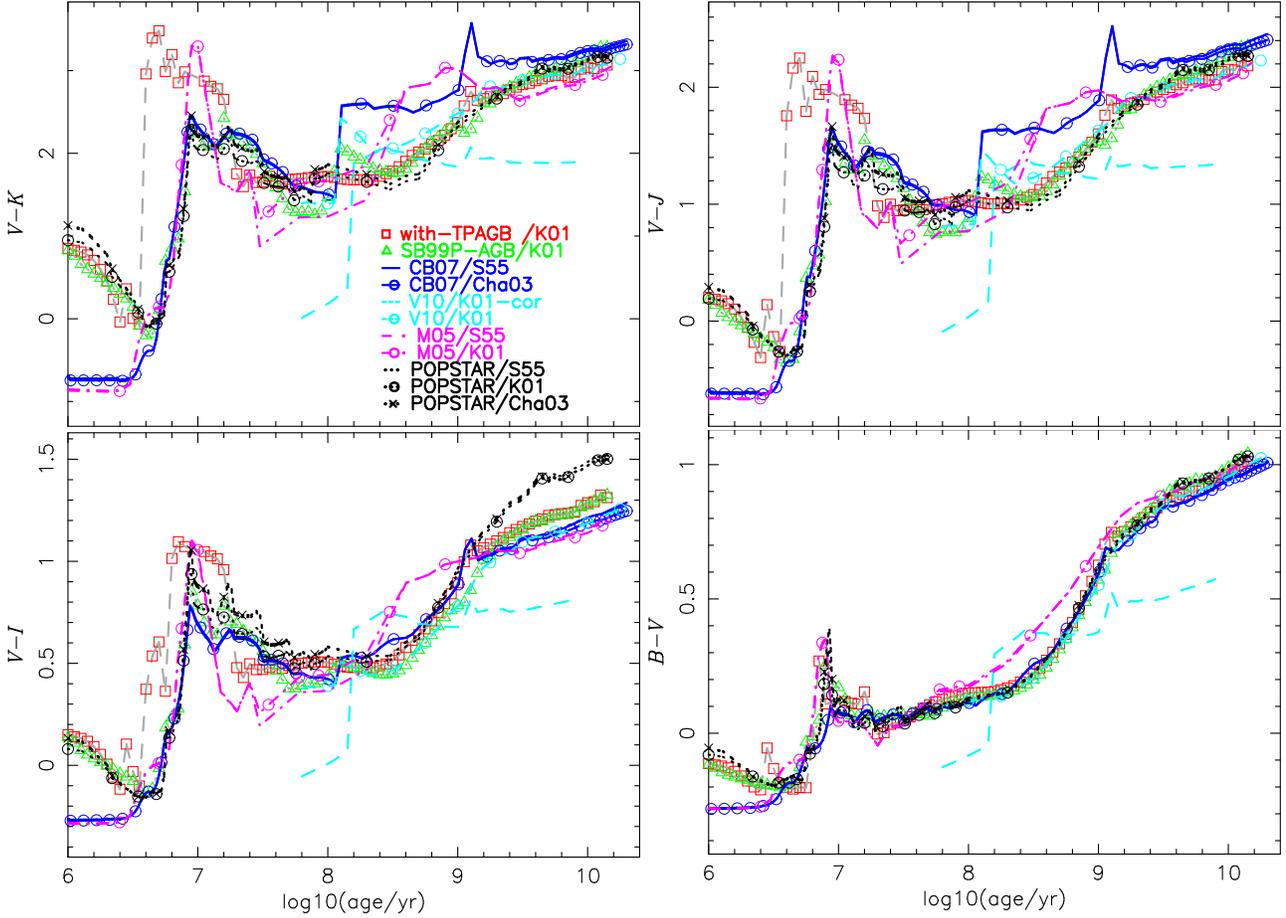

\includegraphics[bb=79 72 504 700,clip,angle=270,scale=.38]{VK-t2-com.ps}
\includegraphics[bb=79 72 504 700,clip,angle=270,scale=.38]{VJ-t3-com.ps}\\
\includegraphics[bb=79 72 560 700,clip,angle=270,scale=.38]{VI-t3-com.ps}
\includegraphics[bb=82 72 560 700,clip,angle=270,scale=.38]{BV-t2-com.ps}
\caption{Comparisons in the $V-K$, $V-J$, $V-I$ and $B-V$ colours among the with-TPAGB-K01 (red open rectangles), SB99P-AGB-K01 (green open triangles), CB07 (blue, solid line for CB07-S55, solid line+circles for CB07-Cha03), V10 (cyan, dashed line for V10-K01-cor, dashed line+circles for V10-K01), M05 (magenta, dot-dashed line for M05-S55, dot-dashed line+circles for M05-K01) and POPSTAR (black, dotted line for POPSTAR-S55, dotted line+circles for POPSTAR-K01, dotted line+crosses for POPSTAR-Cha03) models. }
\label{Fig:ssps-color-com}
\end{figure*}

\begin{table*}
\centering
\caption{Main characteristics of the used EPS ({\hf SB99}, POPSTAR, V10, M05 and CB07) models. The 2nd and 3st columns are for metallicity and age range. The 4th and 5th columns are library of stellar evolution and the name of the corresponding isochrone. The 6th and 7th columns are spectral library and IMFs.}
\begin{tabular}{llll lll}
\hline
    Model     &$Z$& Age & stellar evolution & corresponding  & stellar spectra    & IMF \\
             &   & log($t$/yr)  & library           &   isochrone  & library            &     \\
\hline
  Yunnan-III &0.02&6.0-10.18&  Yunnan-III        & Yunnan-III    & BaSeL             & S55/K01\\
  {\hf SB99} &0.02&6.0-10.18&  Padova-AGB [MG07]& SB99P-AGB    & BaSeL             & S55/K01\\
             &... & ...     &  Padova-STD [G00] & SB99P-STD    & ...               & ...    \\
  POPSTAR    &0.02&5.0-10.18&  Padova-1994+BGS98& BBCFN-1994   & BaSeL+other       & S55/K01/Cha03\\
  V10        &0.02&7.8-10.25&  G00              & Girardi-2002 & MILES             & \ \ \ \ \ /K01/K01-cor\\
  CB07       &0.02&5.0-10.30&  Padova-1994+MG07 & Marigo-2008  & BaSeL-3.1         & S55/\ \ \ \ \ /Cha03 \\
   M05       &0.02&3.5-10.18&  Cassisi + Geneva &              & BaSeL-3.1+LM02    & S55/K01 \\
\hline
\end{tabular}
\label{Tab:moe-rev-used}
\end{table*}

%
In this section we will compare our results with the other studies, which have considered TP-AGB stars in their EPS models. For these models, their main characteristics have been described in Section~\ref{Section:model-ovw}.

Here, we do not use all the models, only use the CB07, M05, V10 and POPSTAR models at solar metallicity. The comparisons in the colours with the {\hf SB99} models have been presented in the Section~\ref{Section:result-com-st99}.
(i) For the CB07, we use the set of models by using the Padova-1994 stellar evolutionary tracks, BaSeL-3.1 stellar spectra library, the S55 and Cha03 IMFs (which we call CB07-S55 and CB07-Cha03).
(ii) For the POPSTAR, we use the set of results by using the S55 with the lower and upper mass limits of 0.15 and 100\,M$_\odot$, K01 and Cha03 IMFs (which we call POPSTAR-S55, POPSTAR-K01 and POPSTAR-Cha03).
(iii) For the V10, we choose the set of models by using the K01 and K01-cor IMFs (which we call V10-K01 and V10-K01-cor).
(iv) For the M05, we use the set of models by using the S55 and K01 IMFs (which we call M05-S55 and M05-K01).
The corresponding ingredients for the used EPS models are given in Table~\ref{Tab:moe-rev-used}. It is worth mentioned that at solar metallicity the M05 and V10 models indeed use the 'Cassisi+Geneva' tracks and G00 isochrones, respectively (in comparison with Table 1).
Moreover, the SB99P-AGB/SB99P-STD isochrones corresponds to the Marigo-2008/Girardi-2002 isochrones, respectively, because the same evolutionary tracks but different construction method are used.

We choose the above EPS models for the following reasons: keep the common ingredients as more as possible. From Table~\ref{Tab:moe-rev-used}, we see that these EPS models have one or several ingredients in common.
This guarantees that the sole effect of ingredient on the results can be derived. M05 has mentioned that EPS models keep the uncertainties inherent in the stellar evolutionary tracks and in the spectral transformations, the isolation of the effect on the final results is important and quantifying the uncertainties is very important since the cosmological inferences that are derived on the basis of galaxy ages and metallicities ultimately rely on the SP models.

\subsection{Comparison in the isochrone}
For the sake of the paper's construct, we first present the comparison in the isochrone among these EPS models before analysing the differences in the results.
The POPSTAR models used the Padova-1994 tracks (together with the revision of the BGS98 isochrone, see Table~\ref{Tab:moe-rev}), from which the BBCFN-1994 isochrones are derived (see Section~\ref{Section:model-ovw}).
At solar metallicity, the V10 models indeed used the G00 isochrones (see Table~\ref{Tab:moe-rev}), which are a part of the Girardi-2002 isochrones (see Section~\ref{Section:model-ovw}).
The Padova-1994 tracks and the prescriptions of MG07 in the used CB07 models (see the first paragraph of Section~\ref{Section:result-oth}) correspond to the Marigo-2008 isochrone (see Table~\ref{Tab:moe-rev} and Section~\ref{Section:model-ovw} for details).
Therefore, in Fig.~\ref{Fig:iso-com}, we also present the BBCFN-1994, Girardi-2002 and Marigo-2008 isochrones at ages of log($t$/yr)= 7, 8, 9 and 10 and $Z$=0.02. These three sets of isochrones are from the Padova's website. For the BBCFN-1994 isochrones, we use $M_{\rm BOL}$=4.83.

\subsection{Comparisons in the colours}
In Fig.~\ref{Fig:ssps-color-com}, we present the $V-K$, $V-J$, $V-I$ and $B-V$ colours for the with-TPAGB-K01, SB99P-AGB-K01, CB07-S55, CB07-Cha03, M05-S55, M05-K01, V10-K01, V10-K01-cor, POPSTAR-S55, POPSTAR-K01 and POPSTAR-Cha03 models.
(i) From the lower right panel of Fig.~\ref{Fig:ssps-color-com}, we see that the $B-V$ colour agrees with each other (except V10-K01-cor) when log($t$/yr)$\ga$6.6.
(ii) From the lower left panel of Fig.~\ref{Fig:ssps-color-com}, we see that the $V-I$ colour is in agreement for all models (except V10-K01-cor) when log($t$/yr)$\ga$6.6, but is not so good as that of $B-V$ colour, the discrepancy in the $V-I$ colour among all models is within $\sim$0.5\,mag. When 8.4$\la$log($t$/yr)$\la$9.0 and log($t$/yr)$\ga$9.0, the $V-I$ colour of M05 and POPSTAR models is greater than the other models, respectively.
(iii) From the top-right panel of Fig.~\ref{Fig:ssps-color-com}, we see that the $V-J$ colour coincides for all models (except V10-K01-cor) when log($t$/yr)$\ga$6.6. But, at small [6.6$\la$log($t$/yr)$\la$7.2] and intermediate [8.0$\la$log($t$/yr)$\la$9.6] ages, the $V-J$ colour of the with-TPAGB-K01 models ($\la$0.6\,mag) and those of the CB07 and M05 models ($\la$0.6\,mag) are significantly greater than the value of the other models, respectively.
(iv) From the top-left panel of Fig.~\ref{Fig:ssps-color-com}, we see that the $V-K$ colour agrees with each other (except V10-K01-cor). However, at small ages [6.6$\la$log($t$/yr)$\la$7.2], the $V-K$ colour of with-TPAGB-K01 and M05 models is greater ($\la$1.0\,mag) than the other models. At intermediate ages [8.0$\la$log($t$/yr)$\la$9.6], the $V-K$ colour of CB07 and M05 models is significantly greater ($\la$1.0\,mag) than the other models.
The following, we will give the comparisons in the colours between ours and other models and analyse the reasons.

\subsubsection {Comparison between with-TPAGB and POPSTAR models:}
The colours of POPSTAR models agree with those of with-TPAGB-K01 models except $V-K$, $V-J$ and $V-I$ in the range of 6.6$\la$log($t$/yr)$\la$7.6. The $V-I$ colour of POPSTAR models is redder than that of with-TPAGB-K01 models at large ages [$\sim$0.2\,mag, log($t$/yr)$\ga$9.0]. Moreover, there is no jump in the evolution of $V-K$ colour (exists for SB99P-AGB models) for POPSTAR models.
The effect of IMF (S55 \& K01 \& Cha03) on the colours [except $V-I$ colour at 7.0$\la$log($t$/yr)$\la$8.0] is insignificant, the differences in the colours between the POPSTAR-S55 and POPSTAR-Cha03 models are small.

POPSTAR-K01 models use the BaSeL+other library and K01 IMF. So the difference in the $V-I$ colour between the POPSTAR-K01 and with-TPAGB-K01 models at large ages is caused by the differences in the stellar evolutionary tracks, spectral library and some techniques in the construction of EPS models.
At large ages, SPs compose of low-mass stars. In these two sets of stellar evolutionary tracks used by the POPSTAR and with-TPAGB-K01 models, the method obtaining TP-AGB tracks (simply synthetic code or complete evolution code), input physics and parameters (such as opacity, EOS and overshooting) are different.
The stellar evolutionary tracks used by the POPSTAR and with-TPAGB-K01 models correspond to the BBCFN-1994 and Yunnan-III isochrones (see Table~\ref{Tab:moe-rev-used}).
From Fig.~\ref{Fig:iso-com}, we see that the first GB along the BBCFN-1994 isochrones has lower temperature than that along the Yunnan-III isochrones at log($t$/yr)=10., this leads to redder $V-I$ colour. Because the first GB does not extend to very low-temperature region, therefore, it does not affect the $V-K$ and $V-J$ colours.

\subsubsection {Comparison between with-TPAGB and V10 models:}
The beginning time of the V10 models is log($t$/yr)$\sim$7.8.
First, from Fig.~\ref{Fig:ssps-color-com}, we see that the effect of IMF (K01 \& K01-cor) on the colours is significant. The results of the V10-K01-cor models are quiet different from those of the V10-K01 and the other EPS models.
The results of the V10-K01 models agree with those of the with-TPAGB-K01 models. However, at intermediate ages [8.0$\la$log($t$/yr)$\la$9.0], the $V-K$ and $V-J$ colours of the V10-K01 models are redder ($\sim$0.4 and 0.3\,mag); and at large ages [log($t$/yr)$\ga$9.0], the $V-I$ colour is bluer ($\sim$0.1\,mag) than the corresponding one of the with-TPAGB-K01 models.

The V10-K01 models use the G00 isochrones, which is a part of the Girardi-2002 isochrones. Moreover, the V10-K01 models use MILES stellar spectra library (MILES comprises some TP-AGB stars although at non-solar metallicity), but the with-TPAGB-K01 models use the BaSeL library.
Therefore, the differences in the colours between the V10-K01 and with-TPAGB-K01 models are mainly caused by the adoptions of different spectral and stellar evolution libraries and different constructional techniques.
Comparing the isochrones in Fig.~\ref{Fig:iso-com}, we see that the TP-AGB stars along the Girardi-2002 isochrones do not extend into the very cool region than those along the Yunnan-III isochrones when 8.0$\la$log($t$/yr)$\la$9.0, and the first GB has lower temperature at large ages, the $V-K$ and $V-J$ colours at intermediate age should be bluer and the $V-I$ colour at large ages should be redder.
But the fact is the contrary. Therefore, the difference in the stellar spectral library (MILES) should be the main factor that causing the differences between the V10-K01 and with-TPAGB-K01 models.

Recall the difference in the $V-K$ and $V-J$ colours between the SB99P-AGB and SB99P-STD models in Fig.~\ref{Fig:ssps-color} and the fact that the SB99P-STD isochrones correspond to the Girardi-2002 isochrones (the V10 and SB99P-STD models use the same stellar evolutionary tracks at intermediate and large ages), we conclude that the $V-K$ and $V-J$ colours will be redder by $\sim$ 0.5\,mag if the V10-K01 models use the Marigo-2008 isochrones, also true for the Yunnan-III isochrones. This is because that the Yunnan-III isochrones are similar to the SB99P-AGB isochrones (conclusion made in Section~\ref{Section:result-com-st99}) and the SB99P-AGB isochrones correspond to the Marigo-2008 isochrones.

\subsubsection {Comparison between with-TPAGB and M05 models:}
First, from Fig.~\ref{Fig:ssps-color-com} we see that the effect of IMF (S55 \& K01) on the colours is insignificant except $V-K$ and $V-J$ colours in the age of 7.6$\la$log($t$/yr)$\la$8.4.
The $V-K$, $V-J$, $V-I$ and $B-V$ colours are redder than the corresponding one of the with-TPAGB-K01 models by $\sim$1\,mag, $\sim$0.6\,mag, $\sim$0.3\,mag and $\sim$0.1\,mag in the age of 8.2$\la$log($t$/yr)$\la$9.0.

The M05 models use the BaSeL library, LM02 spectra for TP-AGB stars and the 'fuel consumption' theorem, therefore, the differences in the colours between the M05-K01 and with-TPAGB-K01 models are partly introduced by the adoption of 'fuel consumption' theorem and the spectra of TP-AGB stars.

\subsubsection {Comparison between with-TPAGB and CB07 models:}
The colours of the CB07 models agree with those of with-TPAGB-K01 models at log($t$/yr)$\ga$6.6 except $V-K$ and $V-J$ colours at intermediate ages [redder by $\sim$1.0\,mag, 8.0$\la$log($t$/yr)$\la$9.6] and $V-I$ colour at large ages [bluer by $\sim$0.2\,mag, log($t$/yr)$\ga$9.0, agrees with V10].
The effect of IMF (S55 \& Cha03) on the colours is insignificant except at large ages [log($t$/yr)$\ga$10.0, a hundredth of a mag] for the CB07 models.

The effect of IMF on the colours is small for both CB07 and with-TPAGB models, and two models give the results with the S55 IMF. CB07 models use BaSeL spectral library and the Padova-1994 tracks together with the description of MG07 (corresponds to the Marigo-2008 isochrones). Therefore,
The differences, in the $V-K$ and $V-J$ colours at intermediate ages and $V-I$ colour at large ages, between the CB07-S55 and with-TPAGB-S55 models are caused by the difference in the stellar evolution tracks and the techniques during the construction of EPS models.
From Fig.~\ref{Fig:iso-com}, we see that TP-AGB stars on the Marigo-2008 isochrones extend into cooler region than those on the Yunnan-III isochrones at intermediate ages, therefore, the redder $V-K$ and $V-J$ colour.
Moreover, from Fig.~\ref{Fig:iso-com}, we indeed see that the technique of building isochrones would be a reason, the Marigo-2008 and SB99P-AGB isochrones use the same evolutionary tracks, but the isochrones are different.

\subsubsection {Comparison among all models:}
Within 6.6$\la$log($t$/yr)$\la$8.0, the SB99P-AGB, CB07 and POPSTAR models (K01-IMF better) are in agreement with each other, the V10 models do not cover this age range.
The main reason for the the agreement is that they use the same stellar evolutionary tracks for massive stars (i.e. the Padova-1994, see Table~\ref{Tab:moe-rev-used}) and stellar spectra library.
The $V-K$ (except for M05), $V-J$ and $V-I$ colours of the with-TPAGB-K01 models are redder than all of the other models. The reason for the differences has been discussed in Section~\ref{Section:result-com-st99}.

At large ages log($t$/yr)$\ga$9.0, all colours of all models (except V10-K01-cor) are in agreement. The CB07 models have reddest $V-K$ and $V-J$ colours; POPSTAR has the reddest $V-I$ colour ($>$ with-TPAGB/SB99P-AGB $>$ CB07/V10 $>$ M05 in turn).

At intermediate ages [8.0$\la$log($t$/yr)$\la$9.0], the $V-K$ and $V-J$ colours of CB07 and M05 models and the $V-I$ colour of M05 model are redder than all the other models. Moreover, the maximal disagreement in the colours among different literatures appears in this range, especially in the $V-J$ and $V-K$ colours.

\subsection{Comparison in the contribution of TP-AGB stars}
In Section~\ref{Section:result}, we conclude that the maximal contribution of TP-AGB stars to the $K$-band is about 50 per cent.
This value is somewhat different from the other authors: \citet{fro90} concluded that the TP-AGB contribution to the total luminosity of SP reaches a maximum of about 40 percent at ages from 1 to 3\,Gyr, falling to less than 10 percent at 10\,Gyr (from M08);

\citet{bru11} mentioned that it can reach to 60 percent in the CB11 models, close to a factor of two more than in the BC03 model; M05 found that 40 percent of the bolometric contribution and 80 percent in $K$ passband.
Part of reasons is that we used the BaSeL library, the spectra of TP-AGB stars are represented by normal M-type giants. Another reason is the less redder TP-AGB stars and shorter during-time are produced when comparing with some EPS models.

\section{Summary and conclusions}
Using the {\hf MESA} stellar evolution code, we compute the consistent stellar models for low-, intermediate-mass and massive stars in the range of 0.01-100\,M$_\odot$. By combining with BaSeL stellar spectra library and the K01 and S55 IMFs, we build the Yunnan-III EPS models, and present colours and ISEDs of solar-metallicity SPs in the range of 1Myr-15\,Gyr.
By comparing the results between Models with-TPAGB and without-TPAGB, we confirm that the inclusion of TP-AGB stars can raise $V-K$, $V-R$ colours and IR flux for SPs at log($t$/yr)$\ga$7.6, the maximal value is at an age of log($t$/yr)$\sim$8.6 (about 0.5-0.2\,mag for colours, about 2 times for $K-$band flux).
Moreover, by comparisons, we find that the colour-evolution trends of Model with-TPAGB are similar to those obtained by the {\hf SB99} code employing the Padova-AGB library.
At last, by comparing the results among ours, CB07, M05, V10 and POPSTAR, we find that the difference in the $B-V$ colour is small, but the difference in the $V-K$ colour is large ($\sim$1\,mag when 8. $\la$log($t$/yr)$\la$9.)
The above conclusion is independent of the adoption of IMF (K01 \& S55 IMFs).

The stellar evolutionary tracks, isochrones, colours and ISEDs of SPs can be obtained on request from the first author or from our website (http://www1.ynao.ac.cn/$\sim$zhangfh/). You can build your EPS models by combining the isochrones with the stellar spectra library and IMF you wanted.
Now the format of stellar evolutionary tracks we provided is the same as that in the {\hf SB99} code, you can input easily them in it to get the SPs' results.
Moreover, the colours involving other passbands or on other systems (for example, HST $F439W-F555W$ colour on AB system) can also be obtained on request.
In this paper we have only presented the results of solar-metallicity SPs - more detailed studies will be given later.

\section*{acknowledgments}
This work was funded by the Chinese Natural Science Foundation (Grant Nos 11273053, 11073049, 11033008, 10821026 \& 2007CB15406), by Yunnan Foundation (Grant No 2011CI053) and by the Chinese Academy of Sciences (KJCX2-YW-T24).
We are also grateful to the referee for suggestions that have improved the quality of this paper.

\bibliography{zfh-mn2}

\appendix

\section{The magnitudes and colours for Model with-TPAGB.}
In this section, we present the integrated magnitudes and colours of solar-metallicity SPs when considering TP-AGB evolution (i.e. Model with-TPAGB). Tables~\ref{Tab:rsts-A} and~\ref{Tab:rsts-B} are the results when using the K01 and S55 IMFs, respectively.

\begin{table*}
\tiny
\centering
\caption{The bolometric magnitude $M_{\rm BOL}$, V-magnitude $M_{\rm V}$, $U-B$, $B-V$, $V-R$, $V-I$, $V-J$, $V-H$ and $V-K$ Colours of solar-metallicity SPs when considering TP-AGB stars (Model with-TPAGB) and using the K01 IMF.}
\begin{tabular}{crr rrr rrr r}
\hline
    log($t$/yr)&$M_{\rm BOL}     $&$M_{\rm V}       $&$U-B       $&$B-V       $&$V-R       $&$V-I       $&$V-J       $&$V-H       $&$V-K       $\\
 \hline
    6.000&$   -17.732$&$   -14.637$&$    -1.252$&$    -0.113$&$     0.073$&$     0.150$&$     0.204$&$     0.399$&$     0.851$\\
     6.050&$   -17.747$&$   -14.654$&$    -1.247$&$    -0.118$&$     0.068$&$     0.140$&$     0.186$&$     0.378$&$     0.827$\\
     6.100&$   -17.770$&$   -14.680$&$    -1.244$&$    -0.122$&$     0.064$&$     0.132$&$     0.173$&$     0.362$&$     0.807$\\
     6.150&$   -17.792$&$   -14.709$&$    -1.238$&$    -0.129$&$     0.057$&$     0.118$&$     0.149$&$     0.333$&$     0.772$\\
     6.200&$   -17.819$&$   -14.745$&$    -1.224$&$    -0.142$&$     0.044$&$     0.092$&$     0.103$&$     0.277$&$     0.704$\\
     6.250&$   -17.849$&$   -14.789$&$    -1.205$&$    -0.159$&$     0.026$&$     0.054$&$     0.034$&$     0.192$&$     0.598$\\
     6.300&$   -17.886$&$   -14.849$&$    -1.180$&$    -0.178$&$     0.003$&$     0.006$&$    -0.054$&$     0.081$&$     0.457$\\
     6.350&$   -17.933$&$   -14.957$&$    -1.142$&$    -0.200$&$    -0.027$&$    -0.058$&$    -0.182$&$    -0.086$&$     0.234$\\
     6.400&$   -17.983$&$   -15.168$&$    -1.085$&$    -0.211$&$    -0.052$&$    -0.117$&$    -0.312$&$    -0.269$&$    -0.039$\\
     6.450&$   -18.030$&$   -16.059$&$    -0.740$&$    -0.055$&$     0.041$&$     0.104$&$     0.140$&$     0.266$&$     0.362$\\
     6.500&$   -17.784$&$   -15.532$&$    -0.845$&$    -0.132$&$    -0.018$&$    -0.031$&$    -0.133$&$    -0.091$&$     0.003$\\
     6.550&$   -17.589$&$   -15.105$&$    -0.960$&$    -0.189$&$    -0.049$&$    -0.101$&$    -0.260$&$    -0.186$&$    -0.062$\\
     6.600&$   -17.437$&$   -14.721$&$    -0.997$&$    -0.201$&$    -0.010$&$     0.373$&$     1.756$&$     2.532$&$     2.960$\\
     6.650&$   -17.285$&$   -14.503$&$    -0.995$&$    -0.208$&$     0.004$&$     0.536$&$     2.164$&$     2.958$&$     3.394$\\
     6.700&$   -17.083$&$   -14.313$&$    -0.983$&$    -0.202$&$     0.018$&$     0.605$&$     2.250$&$     3.050$&$     3.482$\\
     6.750&$   -16.851$&$   -14.233$&$    -0.955$&$    -0.203$&$    -0.020$&$     0.363$&$     1.794$&$     2.564$&$     2.990$\\
     6.800&$   -16.710$&$   -14.253$&$    -0.878$&$    -0.029$&$     0.253$&$     1.014$&$     2.092$&$     2.840$&$     3.194$\\
     6.850&$   -16.563$&$   -14.442$&$    -0.762$&$     0.269$&$     0.431$&$     1.096$&$     1.939$&$     2.571$&$     2.852$\\
     6.900&$   -16.357$&$   -14.261$&$    -0.694$&$     0.109$&$     0.347$&$     1.072$&$     1.981$&$     2.665$&$     2.985$\\
     7.050&$   -15.573$&$   -13.565$&$    -0.682$&$     0.118$&$     0.366$&$     1.048$&$     1.894$&$     2.567$&$     2.873$\\
     7.100&$   -15.385$&$   -13.419$&$    -0.679$&$     0.112$&$     0.365$&$     1.026$&$     1.848$&$     2.512$&$     2.811$\\
     7.150&$   -15.194$&$   -13.251$&$    -0.684$&$     0.109$&$     0.367$&$     1.012$&$     1.821$&$     2.488$&$     2.780$\\
     7.200&$   -15.037$&$   -13.226$&$    -0.635$&$     0.154$&$     0.358$&$     0.959$&$     1.733$&$     2.373$&$     2.653$\\
     7.300&$   -14.685$&$   -13.154$&$    -0.555$&$    -0.003$&$     0.158$&$     0.479$&$     0.985$&$     1.521$&$     1.753$\\
     7.350&$   -14.522$&$   -13.088$&$    -0.516$&$     0.002$&$     0.146$&$     0.430$&$     0.886$&$     1.385$&$     1.597$\\
     7.400&$   -14.367$&$   -12.969$&$    -0.493$&$     0.033$&$     0.176$&$     0.498$&$     1.000$&$     1.519$&$     1.736$\\
     7.450&$   -14.216$&$   -12.898$&$    -0.460$&$     0.038$&$     0.168$&$     0.469$&$     0.944$&$     1.441$&$     1.647$\\
     7.500&$   -14.078$&$   -12.819$&$    -0.433$&$     0.048$&$     0.168$&$     0.470$&$     0.951$&$     1.445$&$     1.653$\\
     7.550&$   -13.943$&$   -12.733$&$    -0.409$&$     0.061$&$     0.172$&$     0.480$&$     0.995$&$     1.506$&$     1.724$\\
     7.600&$   -13.796$&$   -12.652$&$    -0.388$&$     0.077$&$     0.177$&$     0.472$&$     0.953$&$     1.439$&$     1.642$\\
     7.650&$   -13.658$&$   -12.568$&$    -0.371$&$     0.093$&$     0.182$&$     0.475$&$     0.953$&$     1.430$&$     1.630$\\
     7.700&$   -13.530$&$   -12.479$&$    -0.360$&$     0.106$&$     0.186$&$     0.488$&$     0.988$&$     1.476$&$     1.685$\\
     7.750&$   -13.396$&$   -12.389$&$    -0.350$&$     0.117$&$     0.191$&$     0.493$&$     0.995$&$     1.480$&$     1.689$\\
     7.800&$   -13.263$&$   -12.296$&$    -0.341$&$     0.127$&$     0.195$&$     0.500$&$     1.005$&$     1.487$&$     1.695$\\
     7.850&$   -13.134$&$   -12.205$&$    -0.328$&$     0.134$&$     0.197$&$     0.504$&$     1.016$&$     1.498$&$     1.707$\\
     7.900&$   -13.005$&$   -12.119$&$    -0.308$&$     0.140$&$     0.199$&$     0.507$&$     1.022$&$     1.502$&$     1.711$\\
     7.950&$   -12.878$&$   -12.034$&$    -0.285$&$     0.145$&$     0.199$&$     0.507$&$     1.027$&$     1.508$&$     1.716$\\
     8.000&$   -12.750$&$   -11.947$&$    -0.259$&$     0.150$&$     0.197$&$     0.504$&$     1.032$&$     1.513$&$     1.721$\\
     8.050&$   -12.616$&$   -11.861$&$    -0.233$&$     0.151$&$     0.193$&$     0.495$&$     1.022$&$     1.500$&$     1.708$\\
     8.100&$   -12.489$&$   -11.777$&$    -0.207$&$     0.152$&$     0.189$&$     0.487$&$     1.012$&$     1.486$&$     1.692$\\
     8.150&$   -12.361$&$   -11.683$&$    -0.183$&$     0.153$&$     0.187$&$     0.481$&$     1.005$&$     1.477$&$     1.683$\\
     8.200&$   -12.233$&$   -11.582$&$    -0.161$&$     0.156$&$     0.187$&$     0.479$&$     1.003$&$     1.471$&$     1.678$\\
     8.250&$   -12.113$&$   -11.495$&$    -0.133$&$     0.162$&$     0.188$&$     0.482$&$     1.008$&$     1.475$&$     1.683$\\
     8.300&$   -11.997$&$   -11.441$&$    -0.094$&$     0.169$&$     0.186$&$     0.477$&$     1.002$&$     1.466$&$     1.674$\\
     8.350&$   -11.882$&$   -11.372$&$    -0.057$&$     0.177$&$     0.186$&$     0.478$&$     1.006$&$     1.471$&$     1.681$\\
     8.400&$   -11.778$&$   -11.303$&$    -0.018$&$     0.191$&$     0.190$&$     0.488$&$     1.036$&$     1.509$&$     1.722$\\
     8.450&$   -11.670$&$   -11.226$&$     0.020$&$     0.204$&$     0.192$&$     0.497$&$     1.059$&$     1.537$&$     1.754$\\
     8.500&$   -11.565$&$   -11.148$&$     0.053$&$     0.221$&$     0.199$&$     0.514$&$     1.083$&$     1.566$&$     1.785$\\
     8.550&$   -11.466$&$   -11.063$&$     0.081$&$     0.243$&$     0.210$&$     0.541$&$     1.126$&$     1.615$&$     1.838$\\
     8.600&$   -11.374$&$   -10.977$&$     0.107$&$     0.272$&$     0.226$&$     0.577$&$     1.181$&$     1.677$&$     1.904$\\
     8.650&$   -11.284$&$   -10.888$&$     0.131$&$     0.302$&$     0.242$&$     0.612$&$     1.237$&$     1.743$&$     1.974$\\
     8.700&$   -11.190$&$   -10.798$&$     0.150$&$     0.333$&$     0.256$&$     0.640$&$     1.275$&$     1.785$&$     2.016$\\
     8.750&$   -11.095$&$   -10.706$&$     0.164$&$     0.368$&$     0.272$&$     0.669$&$     1.310$&$     1.823$&$     2.054$\\
     8.800&$   -11.011$&$   -10.614$&$     0.169$&$     0.412$&$     0.293$&$     0.708$&$     1.360$&$     1.878$&$     2.110$\\
     8.850&$   -10.926$&$   -10.522$&$     0.168$&$     0.461$&$     0.316$&$     0.748$&$     1.405$&$     1.926$&$     2.156$\\
     8.900&$   -10.856$&$   -10.432$&$     0.163$&$     0.512$&$     0.341$&$     0.792$&$     1.462$&$     1.990$&$     2.220$\\
     8.950&$   -10.800$&$   -10.349$&$     0.163$&$     0.566$&$     0.369$&$     0.840$&$     1.530$&$     2.067$&$     2.297$\\
     9.000&$   -10.773$&$   -10.290$&$     0.170$&$     0.626$&$     0.403$&$     0.896$&$     1.602$&$     2.145$&$     2.373$\\
     9.050&$   -10.839$&$   -10.266$&$     0.202$&$     0.704$&$     0.452$&$     0.998$&$     1.761$&$     2.335$&$     2.574$\\
     9.100&$   -10.807$&$   -10.148$&$     0.217$&$     0.748$&$     0.486$&$     1.080$&$     1.892$&$     2.495$&$     2.748$\\
     9.150&$   -10.508$&$    -9.897$&$     0.186$&$     0.734$&$     0.480$&$     1.051$&$     1.819$&$     2.402$&$     2.642$\\
     9.200&$   -10.364$&$    -9.742$&$     0.185$&$     0.751$&$     0.489$&$     1.067$&$     1.839$&$     2.422$&$     2.662$\\
     9.250&$   -10.238$&$    -9.604$&$     0.194$&$     0.769$&$     0.497$&$     1.084$&$     1.862$&$     2.444$&$     2.685$\\
     9.300&$   -10.114$&$    -9.472$&$     0.204$&$     0.785$&$     0.505$&$     1.100$&$     1.876$&$     2.458$&$     2.697$\\
     9.350&$   -10.013$&$    -9.359$&$     0.220$&$     0.803$&$     0.515$&$     1.117$&$     1.897$&$     2.483$&$     2.721$\\
     9.400&$    -9.954$&$    -9.278$&$     0.242$&$     0.824$&$     0.526$&$     1.140$&$     1.932$&$     2.525$&$     2.766$\\
     9.450&$    -9.870$&$    -9.177$&$     0.257$&$     0.839$&$     0.533$&$     1.156$&$     1.959$&$     2.558$&$     2.801$\\
     9.500&$    -9.784$&$    -9.076$&$     0.273$&$     0.853$&$     0.540$&$     1.171$&$     1.982$&$     2.585$&$     2.830$\\
     9.550&$    -9.699$&$    -8.982$&$     0.288$&$     0.866$&$     0.547$&$     1.183$&$     1.998$&$     2.603$&$     2.848$\\
     9.600&$    -9.618$&$    -8.885$&$     0.303$&$     0.878$&$     0.553$&$     1.197$&$     2.021$&$     2.631$&$     2.878$\\
     9.650&$    -9.528$&$    -8.784$&$     0.315$&$     0.888$&$     0.559$&$     1.208$&$     2.038$&$     2.651$&$     2.900$\\
     9.700&$    -9.438$&$    -8.677$&$     0.331$&$     0.900$&$     0.566$&$     1.225$&$     2.063$&$     2.680$&$     2.931$\\
     9.750&$    -9.304$&$    -8.536$&$     0.335$&$     0.902$&$     0.568$&$     1.229$&$     2.073$&$     2.692$&$     2.945$\\
     9.800&$    -9.189$&$    -8.423$&$     0.349$&$     0.911$&$     0.572$&$     1.234$&$     2.071$&$     2.687$&$     2.938$\\
     9.850&$    -9.068$&$    -8.310$&$     0.363$&$     0.920$&$     0.576$&$     1.235$&$     2.061$&$     2.673$&$     2.920$\\
     9.900&$    -8.963$&$    -8.206$&$     0.383$&$     0.931$&$     0.580$&$     1.241$&$     2.064$&$     2.673$&$     2.919$\\
     9.950&$    -8.938$&$    -8.129$&$     0.418$&$     0.952$&$     0.592$&$     1.274$&$     2.135$&$     2.761$&$     3.018$\\
    10.000&$    -8.853$&$    -8.031$&$     0.447$&$     0.966$&$     0.599$&$     1.287$&$     2.154$&$     2.782$&$     3.040$\\
    10.050&$    -8.747$&$    -7.931$&$     0.480$&$     0.981$&$     0.606$&$     1.292$&$     2.149$&$     2.772$&$     3.026$\\
    10.100&$    -8.709$&$    -7.837$&$     0.516$&$     0.999$&$     0.615$&$     1.325$&$     2.222$&$     2.862$&$     3.128$\\
    10.150&$    -8.578$&$    -7.739$&$     0.552$&$     1.012$&$     0.615$&$     1.312$&$     2.184$&$     2.810$&$     3.066$\\
\hline
\end{tabular}
\label{Tab:rsts-A}
\end{table*}

\begin{table*}
\tiny
\centering
\caption{Similar to Table~\ref{Tab:rsts-A}, but for the results when using the S55 IMF.}
\begin{tabular}{crr rrr rrr r}
\hline
    log($t$/yr)&$M_{\rm BOL}     $&$M_{\rm V}       $&$U-B       $&$B-V       $&$V-R
$&$V-I       $&$V-J       $&$V-H       $&$V-K       $\\
\hline
     6.000&$   -17.235$&$   -14.153$&$    -1.244$&$    -0.115$&$     0.070$&$     0.145$&$     0.194$&$     0.387$&$     0.835$\\
     6.050&$   -17.251$&$   -14.169$&$    -1.239$&$    -0.120$&$     0.065$&$     0.135$&$     0.178$&$     0.367$&$     0.811$\\
     6.100&$   -17.274$&$   -14.195$&$    -1.236$&$    -0.124$&$     0.061$&$     0.127$&$     0.164$&$     0.351$&$     0.792$\\
     6.150&$   -17.295$&$   -14.223$&$    -1.230$&$    -0.130$&$     0.055$&$     0.114$&$     0.141$&$     0.323$&$     0.758$\\
     6.200&$   -17.322$&$   -14.259$&$    -1.217$&$    -0.142$&$     0.042$&$     0.089$&$     0.096$&$     0.269$&$     0.691$\\
     6.250&$   -17.352$&$   -14.302$&$    -1.199$&$    -0.159$&$     0.024$&$     0.051$&$     0.030$&$     0.186$&$     0.589$\\
     6.300&$   -17.388$&$   -14.360$&$    -1.175$&$    -0.177$&$     0.002$&$     0.005$&$    -0.056$&$     0.079$&$     0.452$\\
     6.350&$   -17.434$&$   -14.465$&$    -1.138$&$    -0.199$&$    -0.027$&$    -0.058$&$    -0.180$&$    -0.084$&$     0.234$\\
     6.400&$   -17.484$&$   -14.670$&$    -1.083$&$    -0.210$&$    -0.051$&$    -0.115$&$    -0.308$&$    -0.263$&$    -0.032$\\
     6.450&$   -17.531$&$   -15.541$&$    -0.746$&$    -0.058$&$     0.040$&$     0.101$&$     0.134$&$     0.258$&$     0.356$\\
     6.500&$   -17.295$&$   -15.032$&$    -0.849$&$    -0.133$&$    -0.018$&$    -0.033$&$    -0.136$&$    -0.094$&$     0.002$\\
     6.550&$   -17.108$&$   -14.620$&$    -0.960$&$    -0.189$&$    -0.049$&$    -0.101$&$    -0.261$&$    -0.187$&$    -0.063$\\
     6.600&$   -16.962$&$   -14.250$&$    -0.995$&$    -0.200$&$    -0.011$&$     0.361$&$     1.727$&$     2.500$&$     2.928$\\
     6.650&$   -16.815$&$   -14.041$&$    -0.992$&$    -0.207$&$     0.002$&$     0.521$&$     2.133$&$     2.925$&$     3.361$\\
     6.700&$   -16.620$&$   -13.859$&$    -0.979$&$    -0.201$&$     0.015$&$     0.588$&$     2.220$&$     3.017$&$     3.449$\\
     6.750&$   -16.395$&$   -13.784$&$    -0.952$&$    -0.202$&$    -0.021$&$     0.353$&$     1.767$&$     2.536$&$     2.961$\\
     6.800&$   -16.259$&$   -13.806$&$    -0.876$&$    -0.033$&$     0.246$&$     0.997$&$     2.069$&$     2.815$&$     3.169$\\
     6.850&$   -16.116$&$   -13.994$&$    -0.762$&$     0.260$&$     0.424$&$     1.083$&$     1.922$&$     2.553$&$     2.833$\\
     6.900&$   -15.915$&$   -13.820$&$    -0.695$&$     0.104$&$     0.341$&$     1.059$&$     1.964$&$     2.647$&$     2.966$\\
     7.050&$   -15.150$&$   -13.146$&$    -0.680$&$     0.112$&$     0.359$&$     1.033$&$     1.875$&$     2.546$&$     2.851$\\
     7.100&$   -14.966$&$   -13.006$&$    -0.677$&$     0.107$&$     0.358$&$     1.012$&$     1.828$&$     2.491$&$     2.790$\\
     7.150&$   -14.780$&$   -12.843$&$    -0.681$&$     0.104$&$     0.359$&$     0.997$&$     1.801$&$     2.466$&$     2.758$\\
     7.200&$   -14.626$&$   -12.821$&$    -0.633$&$     0.148$&$     0.351$&$     0.946$&$     1.714$&$     2.353$&$     2.633$\\
     7.300&$   -14.282$&$   -12.753$&$    -0.554$&$    -0.004$&$     0.155$&$     0.473$&$     0.975$&$     1.508$&$     1.739$\\
     7.350&$   -14.122$&$   -12.690$&$    -0.515$&$     0.001$&$     0.144$&$     0.426$&$     0.878$&$     1.374$&$     1.585$\\
     7.400&$   -13.970$&$   -12.575$&$    -0.492$&$     0.032$&$     0.174$&$     0.492$&$     0.990$&$     1.506$&$     1.723$\\
     7.450&$   -13.823$&$   -12.507$&$    -0.460$&$     0.037$&$     0.166$&$     0.464$&$     0.935$&$     1.429$&$     1.635$\\
     7.500&$   -13.687$&$   -12.431$&$    -0.432$&$     0.046$&$     0.166$&$     0.465$&$     0.942$&$     1.434$&$     1.641$\\
     7.550&$   -13.555$&$   -12.348$&$    -0.409$&$     0.060$&$     0.170$&$     0.475$&$     0.986$&$     1.494$&$     1.712$\\
     7.600&$   -13.411$&$   -12.270$&$    -0.388$&$     0.076$&$     0.175$&$     0.468$&$     0.945$&$     1.428$&$     1.630$\\
     7.650&$   -13.276$&$   -12.189$&$    -0.370$&$     0.091$&$     0.180$&$     0.471$&$     0.945$&$     1.420$&$     1.619$\\
     7.700&$   -13.151$&$   -12.103$&$    -0.360$&$     0.104$&$     0.184$&$     0.484$&$     0.980$&$     1.465$&$     1.674$\\
     7.750&$   -13.020$&$   -12.015$&$    -0.349$&$     0.116$&$     0.189$&$     0.489$&$     0.988$&$     1.470$&$     1.678$\\
     7.800&$   -12.889$&$   -11.925$&$    -0.340$&$     0.125$&$     0.193$&$     0.496$&$     0.997$&$     1.477$&$     1.685$\\
     7.850&$   -12.763$&$   -11.836$&$    -0.327$&$     0.132$&$     0.195$&$     0.500$&$     1.009$&$     1.488$&$     1.697$\\
     7.900&$   -12.636$&$   -11.753$&$    -0.307$&$     0.139$&$     0.197$&$     0.504$&$     1.015$&$     1.493$&$     1.701$\\
     7.950&$   -12.512$&$   -11.670$&$    -0.284$&$     0.144$&$     0.197$&$     0.504$&$     1.021$&$     1.499$&$     1.707$\\
     8.000&$   -12.386$&$   -11.586$&$    -0.258$&$     0.148$&$     0.196$&$     0.501$&$     1.026$&$     1.505$&$     1.713$\\
     8.050&$   -12.255$&$   -11.502$&$    -0.232$&$     0.149$&$     0.192$&$     0.493$&$     1.017$&$     1.492$&$     1.699$\\
     8.100&$   -12.131$&$   -11.421$&$    -0.206$&$     0.151$&$     0.188$&$     0.486$&$     1.007$&$     1.479$&$     1.685$\\
     8.150&$   -12.005$&$   -11.329$&$    -0.182$&$     0.153$&$     0.186$&$     0.480$&$     1.001$&$     1.471$&$     1.677$\\
     8.200&$   -11.880$&$   -11.230$&$    -0.160$&$     0.155$&$     0.186$&$     0.479$&$     0.999$&$     1.466$&$     1.672$\\
     8.250&$   -11.761$&$   -11.146$&$    -0.133$&$     0.161$&$     0.188$&$     0.482$&$     1.005$&$     1.471$&$     1.678$\\
     8.300&$   -11.648$&$   -11.093$&$    -0.094$&$     0.169$&$     0.186$&$     0.477$&$     1.000$&$     1.462$&$     1.669$\\
     8.350&$   -11.535$&$   -11.026$&$    -0.057$&$     0.177$&$     0.186$&$     0.479$&$     1.005$&$     1.468$&$     1.677$\\
     8.400&$   -11.433$&$   -10.959$&$    -0.018$&$     0.191$&$     0.190$&$     0.490$&$     1.035$&$     1.506$&$     1.719$\\
     8.450&$   -11.327$&$   -10.885$&$     0.020$&$     0.204$&$     0.193$&$     0.498$&$     1.058$&$     1.535$&$     1.751$\\
     8.500&$   -11.225$&$   -10.808$&$     0.053$&$     0.221$&$     0.199$&$     0.515$&$     1.083$&$     1.564$&$     1.782$\\
     8.550&$   -11.128$&$   -10.725$&$     0.080$&$     0.243$&$     0.211$&$     0.543$&$     1.126$&$     1.613$&$     1.836$\\
     8.600&$   -11.038$&$   -10.641$&$     0.106$&$     0.272$&$     0.227$&$     0.579$&$     1.181$&$     1.676$&$     1.902$\\
     8.650&$   -10.950$&$   -10.554$&$     0.131$&$     0.302$&$     0.243$&$     0.615$&$     1.237$&$     1.741$&$     1.972$\\
     8.700&$   -10.858$&$   -10.466$&$     0.149$&$     0.333$&$     0.257$&$     0.643$&$     1.276$&$     1.784$&$     2.015$\\
     8.750&$   -10.766$&$   -10.375$&$     0.163$&$     0.368$&$     0.273$&$     0.672$&$     1.312$&$     1.823$&$     2.053$\\
     8.800&$   -10.683$&$   -10.285$&$     0.168$&$     0.412$&$     0.294$&$     0.711$&$     1.362$&$     1.879$&$     2.110$\\
     8.850&$   -10.601$&$   -10.195$&$     0.167$&$     0.461$&$     0.317$&$     0.751$&$     1.407$&$     1.927$&$     2.156$\\
     8.900&$   -10.532$&$   -10.107$&$     0.162$&$     0.513$&$     0.343$&$     0.795$&$     1.465$&$     1.991$&$     2.221$\\
     8.950&$   -10.479$&$   -10.026$&$     0.163$&$     0.566$&$     0.370$&$     0.844$&$     1.533$&$     2.068$&$     2.298$\\
     9.000&$   -10.453$&$    -9.968$&$     0.169$&$     0.625$&$     0.404$&$     0.900$&$     1.605$&$     2.147$&$     2.375$\\
     9.050&$   -10.520$&$    -9.946$&$     0.201$&$     0.703$&$     0.453$&$     1.001$&$     1.763$&$     2.335$&$     2.574$\\
     9.100&$   -10.491$&$    -9.830$&$     0.216$&$     0.748$&$     0.488$&$     1.083$&$     1.894$&$     2.495$&$     2.747$\\
     9.150&$   -10.197$&$    -9.583$&$     0.185$&$     0.735$&$     0.482$&$     1.056$&$     1.823$&$     2.405$&$     2.645$\\
     9.200&$   -10.057$&$    -9.431$&$     0.185$&$     0.751$&$     0.491$&$     1.073$&$     1.845$&$     2.426$&$     2.667$\\
     9.250&$    -9.934$&$    -9.296$&$     0.194$&$     0.770$&$     0.500$&$     1.091$&$     1.868$&$     2.450$&$     2.690$\\
     9.300&$    -9.813$&$    -9.166$&$     0.204$&$     0.787$&$     0.508$&$     1.108$&$     1.884$&$     2.465$&$     2.704$\\
     9.350&$    -9.715$&$    -9.055$&$     0.220$&$     0.805$&$     0.518$&$     1.126$&$     1.907$&$     2.491$&$     2.729$\\
     9.400&$    -9.659$&$    -8.977$&$     0.243$&$     0.826$&$     0.530$&$     1.150$&$     1.942$&$     2.534$&$     2.775$\\
     9.450&$    -9.577$&$    -8.877$&$     0.258$&$     0.841$&$     0.537$&$     1.166$&$     1.970$&$     2.566$&$     2.810$\\
     9.500&$    -9.494$&$    -8.779$&$     0.274$&$     0.856$&$     0.545$&$     1.182$&$     1.994$&$     2.595$&$     2.840$\\
     9.550&$    -9.413$&$    -8.687$&$     0.290$&$     0.869$&$     0.552$&$     1.195$&$     2.011$&$     2.614$&$     2.859$\\
     9.600&$    -9.335$&$    -8.593$&$     0.305$&$     0.882$&$     0.558$&$     1.210$&$     2.035$&$     2.642$&$     2.890$\\
     9.650&$    -9.249$&$    -8.494$&$     0.317$&$     0.892$&$     0.565$&$     1.223$&$     2.053$&$     2.663$&$     2.913$\\
     9.700&$    -9.163$&$    -8.390$&$     0.334$&$     0.904$&$     0.573$&$     1.240$&$     2.079$&$     2.694$&$     2.946$\\
     9.750&$    -9.034$&$    -8.252$&$     0.339$&$     0.907$&$     0.575$&$     1.247$&$     2.091$&$     2.707$&$     2.961$\\
     9.800&$    -8.923$&$    -8.142$&$     0.353$&$     0.917$&$     0.581$&$     1.254$&$     2.091$&$     2.705$&$     2.956$\\
     9.850&$    -8.807$&$    -8.033$&$     0.368$&$     0.926$&$     0.585$&$     1.257$&$     2.085$&$     2.693$&$     2.942$\\
     9.900&$    -8.707$&$    -7.931$&$     0.388$&$     0.938$&$     0.590$&$     1.265$&$     2.090$&$     2.696$&$     2.943$\\
     9.950&$    -8.685$&$    -7.857$&$     0.424$&$     0.959$&$     0.602$&$     1.299$&$     2.160$&$     2.782$&$     3.039$\\
    10.000&$    -8.604$&$    -7.763$&$     0.454$&$     0.974$&$     0.610$&$     1.314$&$     2.180$&$     2.804$&$     3.063$\\
    10.050&$    -8.504$&$    -7.666$&$     0.487$&$     0.990$&$     0.618$&$     1.321$&$     2.178$&$     2.797$&$     3.052$\\
    10.100&$    -8.469$&$    -7.576$&$     0.523$&$     1.008$&$     0.628$&$     1.355$&$     2.251$&$     2.885$&$     3.152$\\
    10.150&$    -8.345$&$    -7.482$&$     0.560$&$     1.022$&$     0.629$&$     1.345$&$     2.217$&$     2.838$&$     3.096$\\
\hline
\end{tabular}
\label{Tab:rsts-B}
\end{table*}

\section{Evolutionary results of TP-AGB stars}
\label{Section:appen-TPAGB}
\begin{figure}
\includegraphics[angle=270,scale=.350]{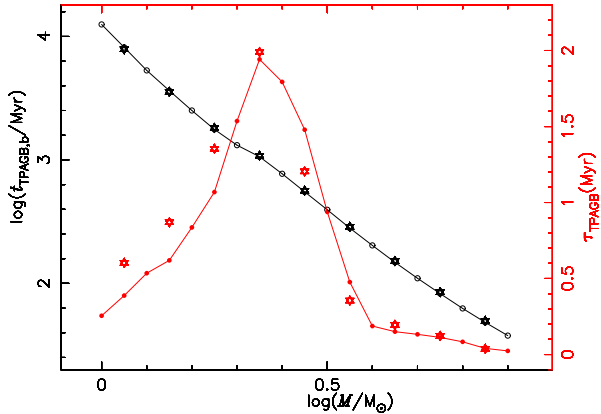}
\caption{The onset time [log($t_{\rm TPAGB,b}$), left y-axis] and during [$\tau_{\rm TPAGB}$, right y-axis] of TP-AGB phase as a function of stellar mass. The results, obtain by using the standard or the C13 set of input parameters and physics in the {\hf MESA} code, are represented by Line+circles and stars, respectively.}
\label{Fig:during-tp}
\end{figure}

\begin{figure}
\includegraphics[angle=270,scale=.350]{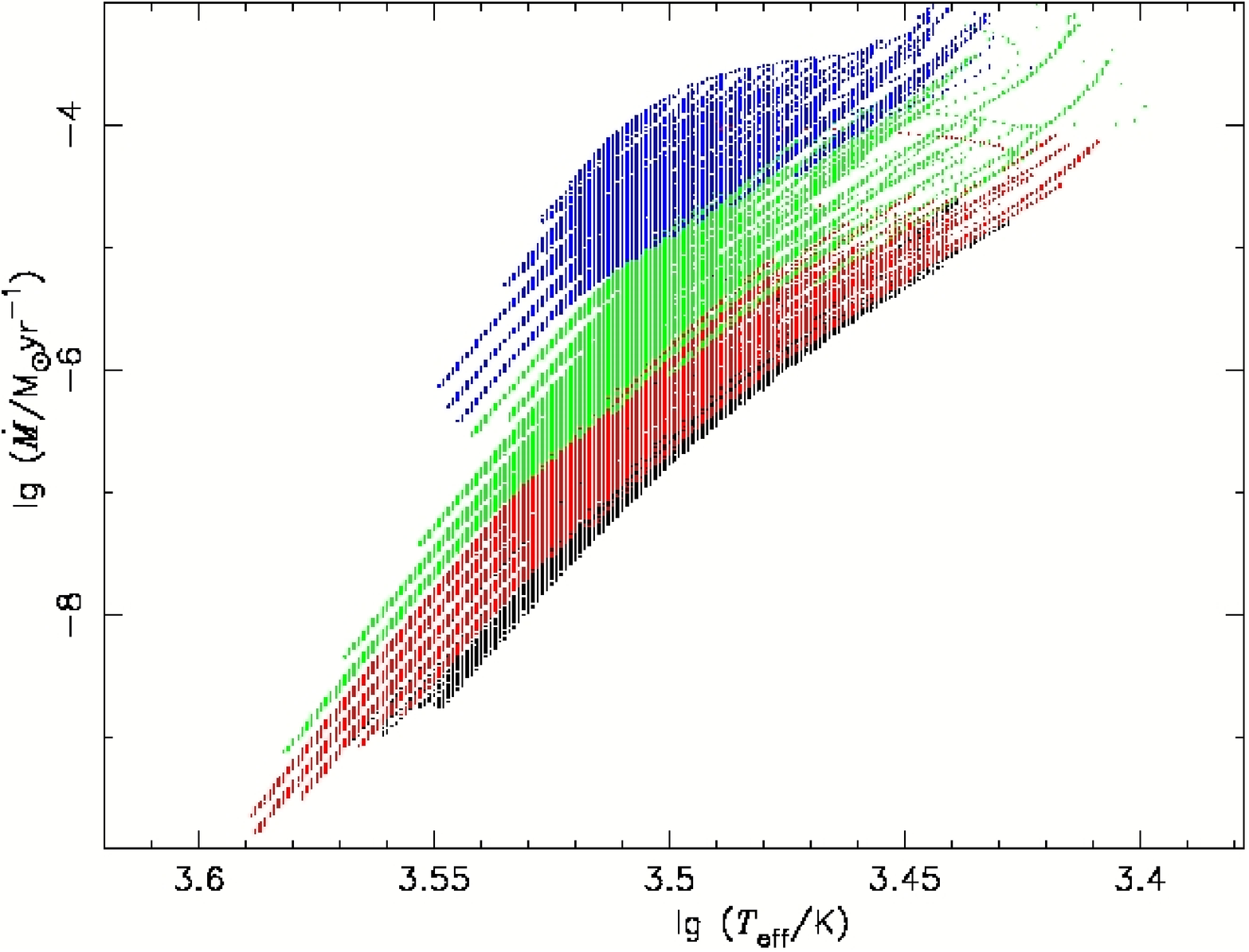}
\caption{The relation between $T_{\rm eff}$ and mass loss $\dot M$ for TP-AGB phase of stars with $1.0 \la M/{\rm M}_\odot \la 7.9$. The symbols have the same meaning as in Fig.~\ref{Fig:tp-hrd}. These results are obtained by using the standard set of input parameters and physics in the {\hf MESA} code.}
\label{Fig:tp-wind}
\end{figure}

\begin{table}
\centering
\caption{The onset time $t_{\rm TPAGB,b}$ (the 2nd column), duration $\tau_{\rm TPAGB}$ (the 3st column) and the pulsation time $n_{\rm TPAGB}$ (the 4th column) of TP-AGB phase for solar-metallicity stars with different masses (the 1st column). Top panel is for the standard set of parameters and physics in the {\hf MESA} code. Bottom panel is for the C13 set of parameters and physics (which is described in Appendix~\ref{Section:appen-TPAGB-2}).}
\begin{tabular}{lcc r}
\hline
  Mass     &  $t_{\rm TPAGB, b}$  & $\tau_{\rm TPAGB}$ & $n_{\rm TPAGB}$  \\
  (M$_\odot$) &  (Myr)            & (Myr)           &               \\
\hline
 & & standard set & \\
\hline
      1.00&12510.3662&    0.2549&     4\\
      1.12& 8157.0493&    0.3877&     6\\
      1.26& 5308.0205&    0.5347&     8\\
      1.41& 3636.1863&    0.6194&     8\\
      1.58& 2512.2832&    0.8362&    11\\

      1.78& 1766.8214&    1.0685&    14\\
      2.00& 1315.7844&    1.5359&    19\\
      2.24& 1060.6736&    1.9417&    25\\
      2.51&  772.4565&    1.7934&    26\\
      2.82&  551.9109&    1.4787&    25\\

      3.16&  394.5910&    0.9380&    22\\
      3.55&  280.5102&    0.4766&    19\\
      3.98&  203.0824&    0.1873&    16\\
      4.47&  148.1105&    0.1500&    18\\
      5.01&  109.9674&    0.1325&    21\\

      5.62&   82.6705&    0.1124&    23\\
      6.31&   62.7118&    0.0836&    23\\
      7.08&   48.3612&    0.0402&    18\\
      7.94&   37.7756&    0.0235&    14\\
\hline
 & & C13 set & \\
\hline
      1.12& 7854.9492&    0.6035&     8\\
      1.41& 3556.3940&    0.8694&    11\\
      1.78& 1814.3972&    1.3538&    16\\
      2.24& 1081.0680&    1.9901&    25\\
      2.82&  564.3586&    1.2044&    23\\
      3.55&  287.4057&    0.3549&    18\\
      4.47&  151.7789&    0.1945&     6\\
      5.62&   84.9745&    0.1219&     6\\
      7.08&   49.7702&    0.0379&    14\\

\hline
\end{tabular}
\label{Tab:tpagb}
\end{table}

In this section, we will present the evolutionary results of TP-AGB stars.
In Appendix~\ref{Section:appen-TPAGB-1}, we present the onset time, lifetime and inter-pulse time of TP-AGB phase, the maximal mass of star experiencing TP-AGB phase, and some comparisons with those of MG07. In Appendix~\ref{Section:appen-TPAGB-1}, all results are obtained by using the set of input physics and parameters, which have been described in Section~\ref{Section:model-des} and called the standard set, in the {\hf MESA} code.
In Appendix~\ref{Section:appen-TPAGB-2}, we present the results by using another set of input parameters and physics used in the {\hf MESA} code, and discuss the effects on the onset time and lifetime of TP-AGB phase.

\subsection{Some results by using the standard set of physics and parameters}
\label{Section:appen-TPAGB-1}

\subsubsection{Onset time and duration of TP-AGB phase:}
First, in Table~\ref{Tab:tpagb} we present the onset time ($t_{\rm TPAGB,b}$, accurately speaking, it is the average time during the first thermal pulse $t'_{1}$), the pulsation time ($n_{\rm TPAGB}$) and the duration or lifetime ($\tau_{\rm TPAGB}$, the difference between the average time during the last thermal pulse $t'_{n_{\rm TPAGB}}$ and $t'_{1}$) of TP-AGB phase as a function of stellar mass.
Then, in Fig.~\ref{Fig:during-tp} we give log($t_{\rm TPAGB,b}$) and $\tau_{\rm TPAGB}$ as a function of stellar mass, which is in the range of 1.0-7.9M$_\odot$.

From them, we see that the logarithmic onset-time of TP-AGB phase [log($t_{\rm TPAGB,b}$] linearly decreases with the logarithmic initial-mass log($M$).
The lifetime of TP-AGB phase reaches the maximum ($\sim 2$\,Myr) at $M$=2.24M$_\odot$. This differs from that of MG07: the maximal during of $\sim 4$\,Myr at $M$=2.M$_\odot$ for M-type TP-AGB stars (see the left-top panel of their Fig.\,22).
That is to say, the maximal lifetime of TP-AGB phase is half of the MG07 result.

\subsubsection{the maximal mass of stars experiencing TP-AGB phase:}
Here need to be mentioned is the maximal mass of stars experiencing TP-AGB phase ($M_{\rm max, TPAGB}$). MG07 have obtained that the upper mass limit for the development of the AGB phase is located close to 5\,M$_\odot$ for all values of $Z$ as a consequence of overshooting.
This value of 5\,M$_\odot$ is smaller than that of 8\,M$_\odot$ found in earlier models (see the review paper of \citealt{van03} and \citealt{her05}). Moreover, the classical results found an important TP-AGB contribution at 5\,M$_\odot$ in overshooting tracks (Padova) and at 3\,M$_\odot$ in tracks without overshooting (M98).
Therefore, 5\,M$_\odot$ is the highest initial mass to be considered in the MG07 work, but they also mentioned that the value of $M_{\rm max, TPAGB}$ is slightly uncertain because the initial phase of carbon burning has not been followed in detail.
In the FSPS models (described in Section~\ref{Section:model-ovw}), it is cited from M08 that $M_{\rm max, TPAGB}$ is 5\,M$_\odot$.

In our calculations, $M_{\rm max, TPAGB}$ is greater than 8\,M$_\odot$ at solar metallicity. In the Fig. 22 of \citet{pax11}, they presented a grid of MESA star evolutionary tracks with mass ranging from 2 to 10\,M$_\odot$ with $Z$=0.02, and in their paper they said that the 8 and 10\,M$_\odot$ models start to ignite carbon burning off center, this implies that these stars are experiencing envelope burning which keeps the TP-AGB contribution low, where 2-7\,M$_\odot$ models do not and will presumably go on form C/O core white dwarfs.
Therefore, the $M_{\rm max, TPAGB}$ is greater than 8\,M$_\odot$, and for stars with $M>$8\,M$_\odot$, the timescale on TP-AGB phase is very short and the contribution is smaller.

\subsubsection{the minimal temperature for superwind:}
In Fig.~\ref{Fig:tp-wind}, we present the TP-AGB stars on the $T_{\rm eff}$ and mass-loss $\dot{M}$ plane, from it we see that the minimal temperature for superwind is log($T_{\rm eff}$/K)$\sim$3.5.

\subsection{Results by using another set of physics and parameters}
\label{Section:appen-TPAGB-2}
In this work, we only calculate the stellar evolution models and present the SP's results at solar metallicity. This can not prove the reliability of the standard set of physics and parameters, in the {\hf MESA} code, on the TP-AGB evolution. Therefore, we change some input parameters and physics in the standard set by mimicing the set used by \citet{pax11} to explain the formation of a C13 pocket ($M=$2\,M$_\odot$ and $Z$=0.01). This set of physics and parameters is called the C13 set.

In the C13 set, the changed parameters and physics are as follows. (i) Overshooting: at both boundaries for H-burning and non-burning convective zones, $f$ (see Section 2.1) is changed from 0.01 to 0.014. Moreover, overshooting is allowed for He-burning and Z-burning zones, and at both boundaries $f$=0.014. During the TDU $f$=0.126 at the bottom of the convective envelope and $f$=0.008 at the bottom of the H-shell flash convection zone. (ii) Opacity: CO enhanced opacity is used. (iii) Mass loss: the coefficient of \citet{blo95} mass-loss-law $\eta_{\rm B}$ is changed from 0.1 to 0.05. (iv) Atmosphere: the usage of simple photosphere model is changed to the usage of model atmosphere tables for photosphere.
Similarly, we use the C13 set of input parameters and physics to obtain the luminosity of 1\,M$_\odot$ star at an age of 4.57$\times$10$^9$yr, log($L/{\rm L}_\odot$)=0.012 for the GS98 opacity while log($L/{\rm L}_\odot$)=0.012 for the GN93 opacity. These luminosities are within one hundredth of a magnitude.

Using the C13 set of physics and parameter in the {\hf MESA} code, we obtain the onset time $t_{\rm TPAGB,b}$, lifetime $\tau_{\rm TPAGB}$ and inter-pulse time $n_{\rm TPAGB}$ of TP-AGB phase. These results are given in the bottom panel of Table~\ref{Tab:tpagb}. Moreover, in Fig.~\ref{Fig:during-tp} we also plot log($t_{\rm TPAGB,b}$) and $\tau_{\rm TPAGB}$ by using the C13 set in the {\hf MESA} code.
By comparisons with the results of the standard set, we found that the variation of $t_{\rm TPAGB,b}$ is little, the $\tau_{\rm TPAGB}$ of low-mass stars becomes to be greater (by $\sim$0.2\,dex) and $\tau_{\rm TPAGB}$ of intermediate-mass stars smaller (no more than 0.05\,dex) than the corresponding one,
and the maximal lifetime, obtained by using the C13 set in the {\hf MESA} code, can not reach to the value of MG07 (the maximum of $\sim4$\,Myr).
The input physics and parameters in the work of MG07 are very differ from those in the {\hf MESA} code. In the following we describe the parameters used by MG07.
\begin{itemize}
\item
{\bf Opacity:} they use variable molecular opacity by means of the \citet[][hereafter M02]{mar02} method, i.e. opacities are consistent with the changing photospheric chemical composition of AGB envelope. The M02 method is different from the commonly used one, i.e. interpolations are preformed as a function of temperature, density and hydrogen content for a fixed initial metallicity (MG07).

\item
{\bf Mass loss}: MG07 use formalisms for the mass-loss rates derived from pulsating dust-driven wind models of C- and O-rich AGB stars, and the value of mass-loss depends on whether a stellar model is oxygen-rich or carbon-rich.

\item {\bf Overshooting}:
MG07 use two parameters: $\Lambda_c$ and $\Lambda_e$. $\Lambda_c$ describes the extent across the border of the convective zone and $\Lambda_e$ is the overshooting at the lower boundary of convective envelope. Both are expressed in units of pressure scale height.
When $M$$\le$1M$_\odot$, $M$$\ge$1.5M$_\odot$ and 1$\le$$M/\rm M_\odot$$\le$1.5, $\Lambda_c$ is 0., 0.5 and $M$-1.0, respectively. In the stage of core helium burning $\Lambda_c=0.5$.
When $M$$\le$0.6M$_\odot$, 0.6$\le$$M/\rm M_\odot$$\le$2.0, 2.0$\le$$M/\rm M_\odot$$\le$2.5 and $M\ge$2.5M$_\odot$,  $\Lambda_e$ is 0., 0.25, 0.25-0.7 and 0.7.

\item  {\bf Temperature:}  the $T_{\rm eff}$ of TP-AGB models is derived with the aid of complete integration of static envelope models, extending from the photosphere down to the core (MG07).
\end{itemize}

The different choice of all these parameter (opacity, the mass loss, overshooting, atmosphere) would alter or affect the evolution of TP-AGB stars.
For example, the adoption of mass-loss law would affect TP-AGB evolution (M05).
Therefore, there exists disagreement between the MG07 and our models.

Similarly, in the 'fuel consumption' technique, uncertainties in mass loss, mixing and efficiency of hydrogen burning at the bottom of the convective envelope together also affect the TP-AGB evolutions (see M05).
These disagreement further affects the SP's results with 'fuel consumption' technique.
M98 has made a conclusion that the adoption of the \citet{blo91} mass-loss law would make the models luminous and evolution faster than that predicted by the standard core mass-luminosity relation for 7\,M$_{\odot}$ star, resulting drastic decrease of the TP-AGB lifetime and smaller contribution of TP-AGB phase to the ISED.

\bsp
\label{lastpage}
\end{document}